\documentclass[aps,prd,10pt,twocolumn,preprintnumbers,floatfix,nofootinbib]{revtex4-1}
\pdfoutput=1
\usepackage{graphicx}
\usepackage{bm}
\usepackage{times}
\usepackage{slashed}
\usepackage[usenames,dvipsnames,svgnames,table]{xcolor}
\usepackage[normalem]{ulem}
\usepackage{lipsum}
\usepackage{subfigure}
\usepackage{amsfonts}		
\usepackage{amsmath}		
\usepackage{amssymb}		
\newcommand{\be}{\begin{equation}}
\newcommand{\ee}{\end{equation}}
\newcommand{\bea}{\begin{eqnarray}}
\newcommand{\eea}{\end{eqnarray}}
\newcommand{\of}[1]{\left( #1 \right)}

\newcommand{\MeV}{~\mathrm{MeV}}
\newcommand{\gev}{~\mathrm{GeV}}
\newcommand{\tev}{~\mathrm{TeV}}
\newcommand{\invfb}{~\mathrm{fb}^{-1}}

\newcommand{\CL}[1]{#1\%~\mathrm{C.L.}}
\newcommand{\order}[1]{\mathcal{O}(#1)}

\newcommand{\neut}[1]{\tilde{\chi}_#1^0}
\newcommand{\chargpm}[1]{\tilde{\chi}_#1^\pm}
\newcommand{\chargp}[1]{\tilde{\chi}_#1^+}

\newcommand{\mneut}[1]{m_{\tilde{\chi}_#1^0}}
\newcommand{\mcharg}[1]{m_{\tilde{\chi}_#1^\pm}}

\newcommand{\ifb}{~\mathrm{fb}^{-1}}

\newcommand{\IceCube}{\textsc{IceCube}}
\newcommand{\LEP}{\textsc{LEP}}
\newcommand{\LHC}{\textsc{LHC}}

\newcommand{\CMS}{\textsc{CMS}}
\newcommand{\XenonT}{\textsc{Xenon1T}}
\newcommand{\XenonH}{\textsc{Xenon100}}
\newcommand{\PandaX}{\textsc{PandaX}}
\newcommand{\LUX}{\textsc{Lux}}
\newcommand{\FermiLAT}{\textsc{Fermi-LAT}}

\newcommand{\fb}{~\mathrm{fb}}
\newcommand{\etmiss}{\slashed{E}_T}
\begin{document}

\title{The Not-So-Well Tempered Neutralino}

\author{Stefano Profumo}
\email{profumo@ucsc.edu}
\author{Tim Stefaniak}
\email{tistefan@ucsc.edu}
\author{Laurel Stephenson-Haskins} 
\email{lestephe@ucsc.edu}

\preprint{SCIPP 17/08}

\affiliation{Department of Physics and Santa Cruz Institute for Particle Physics, University of California, Santa Cruz, CA 95064, USA}

\begin{abstract}
Light electroweakinos, the neutral and charged fermionic supersymmetric partners of the Standard Model $\mathrm{SU}(2)\times\mathrm{U}(1)$ gauge bosons and of the two $\mathrm{SU}(2)$ Higgs doublets, are an important target for searches for new physics with the Large Hadron Collider (\LHC). However, if the lightest neutralino is the dark matter, constraints from direct dark matter detection experiments rule out large swaths of the parameter space accessible to the LHC, including in large part the so-called ``well-tempered'' neutralinos. We focus on the Minimal Supersymmetric Standard Model (MSSM) and explore in detail which regions of parameter space are not excluded by null results from direct dark matter detection, assuming exclusive thermal production of neutralinos in the early universe, and illustrate the complementarity with current and future LHC searches for electroweak gauginos. We consider both bino-Higgsino and bino-wino ``not-so-well-tempered'' neutralinos, i.e.\ we include models where the lightest neutralino constitutes only part of the cosmological dark matter, with the consequent suppression of the constraints from direct and indirect dark matter searches.
\end{abstract}

\maketitle

\section{Introduction}

The ``well-tempered neutralino'', an expression coined in Ref.~\cite{ArkaniHamed:2006mb}, emerges from the notion that \LEP\ results \cite{lep}  forced supersymmetric particle masses to a range where the thermal dark matter (DM) relic density is no longer ``natural'', but rather, is ``critically'' tuned (to use the language of Ref.~\cite{ArkaniHamed:2006mb}), i.e.\ it results from some fine-tuning of the relevant parameters, including soft supersymmetry breaking masses, in the theory. Ref.~\cite{ArkaniHamed:2006mb} then argues that this is somewhat similar to the naturalness problem associated with electroweak symmetry breaking, where, after \LEP, parameters needed to be somewhat tuned to reproduce the correct mass of the $Z$ boson. The well-tempered neutralino  corresponds to the particular fine tuning of the mass parameters relevant to the neutralino sector: the soft-supersymmetry breaking masses associated with the fermionic partners of the $\mathrm{U}(1)_Y$ and $\mathrm{SU}(2)_L$ gauge bosons,  $M_1$ and $M_2$, respectively, and the Higgsino mass parameter, $\mu$. As a result, well-tempered neutralinos correspond to scenarios where $M_{1}$ or $M_{2} \sim \mu$ (bino-Higgsino or wino-Higgsino mixed~\cite{ref16,ref24,Masiero:2004ft,ref22,ref23}) or $M_1\sim M_2$ (bino-wino mixed~\cite{Masiero:2004ft,Baer:2005zc,Baer:2005jq,ref27}). Similar scenarios of finely-tuned neutralino admixtures were explored prior to Ref.~\cite{ArkaniHamed:2006mb}, see for instance Ref.~\cite{ref24,Masiero:2004ft}.

In the \LHC\ era, the situation pictured above has not substantially changed --- rather, it has possibly become more dire, as a result of null searches for squarks, gluinos and, as we will discuss extensively here, for electroweakinos (by which we mean neutralinos and charginos). Additionally, and perhaps more importantly, the decade after Ref.~\cite{ArkaniHamed:2006mb} appeared brought a substantial improvement in the limits from {\em direct} dark matter searches by over three orders of magnitude in the DM-nucleon scattering cross section, severely impacting the mixed gaugino-Higgsino scenarios producing the correct thermal relic neutralino density. This has been pointed out long ago, e.g., in Ref.~\cite{Baer:2006te}, which argued that there are veritable ``target'' direct detection rates for well-tempered neutralinos. There exist caveats to this statement, however: first, in the limit of null Higgsino fraction, that could be realized for a well-tempered bino-wino scenario (see Sec.~\ref{sec:binowino}), or, similarly, in the limit of null gaugino fraction, spin-independent and spin-dependent direct detection rates vanish, since the coupling to Higgs bosons is proportional to the Higgsino-gaugino mixing, and the coupling to the $Z$ boson is proportional to the Higgsino fraction (but also vanishes for pure Higgs\-inos, which are Dirac fermions); secondly, it was realized that there exist {\em blind spots}, where combinations of parameters conjure to cancel the lightest neutralino couplings to the $Z$ and/or to the Higgs boson(s)~\cite{Cheung:2012qy}. The blind spots arise only for particular relative signs of $M_1,\ M_2$ and $\mu$, and depend on the value of the ratio of the two Higgs doublets' vacuum expectation values (vev), $\tan\beta$ \cite{Cheung:2012qy}. The study of blind spots was generalized to simplified models in Ref.~\cite{Cheung:2013dua} in the Higgs decoupling limit, where all non-standard Higgs bosons of the Higgs sector of the Minimal Supersymmetric Standard Model (MSSM) are heavy. Additionally, blind spots for spin-independent searches can also occur via interferences between the two CP-even Higgs bosons $h$ and $H$, for certain relative values (depending on conventions) of the Higgsino mass parameter, $\mu$, over the gaugino masses, $M_{1,2}$, as first noticed in Ref.~\cite{Huang:2014xua}. 

Ref.~\cite{Han:2016qtc} studied the complementarity of dark matter searches in the case of blind spots, pointing out three important facts: (\emph{i}) a blind spot for spin-independent direct detection is not blind for spin-dependent, as well as for searches for neutrinos from the Sun with \IceCube\ (whose rate depends on the capture rate of neutralinos in the Sun, primarily driven by spin-dependent interactions); (\emph{ii}) blind spots typically correspond to models where the predicted gamma-ray signal is {\em too low} to be detectable with the \textsc{Fermi Large Area Telescope (LAT)}; and (\emph{iii}) searches for supersymmetry with the \LHC\ can significantly cover blind spots.

Ref.~\cite{Farina:2011bh} pointed out in 2011 that the (then) recent \XenonH\ results \cite{Aprile:2011hi} were in tension with a rather large portion of the well-tempered bino-Higgsino scenario, barring resonant annihilation channels via the light Higgs boson $h$ or the $Z$ boson, and that the entire parameter space could be ruled out with an increment of a factor of a few in the sensitivity of direct dark matter searches. Indeed, Ref.~\cite{Baer2016} argues that the entire well-tempered neutralino scenario is ruled out by direct detection exclusion limits from \XenonH~\cite{Aprile:2011hi}, \PandaX~\cite{Tan:2016zwf}, and \LUX~\cite{Akerib:2016vxi};
however, Ref.~\cite{Baer2016} denotes with ``well-tempered neutralino'' essentially only bino-Higgsino mixed scenarios, and does not consider bino-wino neutralino admixtures or pure states. Similar conclusions are reached in Ref.~\cite{Badziak:2017the}, which illustrates how a bino-dominated bino-Higgsino mixed state is ruled out by the \LUX\ constraints unless $\tan\beta\lesssim 3$, assuming the Higgs decoupling limit; this relatively low $\tan\beta$ value, in combination with the measurement of the Higgs mass, then forces scalar top (stop) partners to be heavier than $\sim25\tev$. Otherwise, non-standard Higgs bosons need to be light, i.e.\ below around $400\gev$, and $\tan\beta\sim10$ in order for blind spots to appear. Ref.~\cite{Badziak:2017the} also derives a lower limit to the lightest neutralino mass in the Higgsino-like bino-Higgsino mixed case, which again depends on the mass of the ``heavy'' non-standard Higgs bosons.


A similar study of mixed bino-Higgsino scenarios was featured in Ref.~\cite{1701.02737}, which specifically addressed direct detection blind spots arising from interference in the CP-even Higgs boson exchange diagrams relevant for spin-independent searches. In this context Ref.~\cite{1701.02737} addressed, along similar lines to Ref.~\cite{Han:2016qtc}, how a ``holistic'' approach including direct and indirect searches and collider searches for both heavy Higgs boson and electroweakinos could test the remaining open sections of parameter space. More studies of the interplay of dark matter experiments and LHC searches in the mixed bino-Higgsino dark matter scenario can be found in Refs.~\cite{Barducci:2015ffa,Abdughani:2017dqs}.

%

Traditionally, experimental searches at LEP and the LHC for neutralinos and charginos have hinged upon a same-sign dilepton or multilepton signature associated with missing transverse energy, $\etmiss$, assuming the direct production of a heavier neutralino and the lightest chargino, or pairs of heavier neutralinos or charginos. The final state leptons can either arise from cascade or three-body decays of the neutralinos and charginos via [virtual] scalar leptons (sleptons), or in neutralino and chargino decays into the lightest neutralino --- the lightest supersymmetric particle (LSP) --- and a $W$ or $Z$ boson, which can subsequently decay leptonically. \LHC\ searches for this signature~\cite{ATLAS:2016uwq,CMS:2016gvu,CMS:2017fdz} are competitive over large swaths of the gaugino mass parameter space, currently excluding lightest chargino/second neutralino mass values of up to $\sim 450\gev$ (in a simplified scenario where the Higgs\-ino mass parameter, $\mu$, as well as squarks and sleptons are decoupled). However, these searches become insensitive in scenarios with small mass differences $\Delta m$ between the decaying charginos/neutralinos and the LSP, in which case the 4-momenta of the final state leptons are typically small due to phase space suppression, and these leptons escape detection. Indeed, as we will discuss in more detail in this work, well-tempered neutralino dark matter scenarios quite generically feature such a compressed electroweakino mass spectrum, and thus traditional multilepton + $\etmiss$ collider searches are rather insensitive.


More recently, dedicated \LHC\ searches for scenarios with a compressed electroweakino mass spectrum have been performed, requiring two final state leptons, $\etmiss$, and an additional initial state radiated (ISR) jet. The additional ISR jet recoils against the remaining objects, thus giving rise to larger transverse momenta of the two leptons. These searches have so far only been carried out by the \CMS\ experiment. We shall include the latest results from these searches as well as from the traditional multilepton + $\etmiss$ search by the \CMS\ collaboration in our work.

In addition to the search strategies outlined above, Ref.~\cite{Bramante:2014dza} addressed and proposed a different search path based on the possibility of radiative neutralino decays into the LSP and a photon, potentially giving rise to an opposite-sign dilepton signature associated with $\etmiss$ and a detectable photon.

The focus of the present analysis is on what we refer to as \emph{not-so-well tempered} neutralinos: by that we mean  both the well-tempered case, where the thermal relic abundance matches the observed cosmological dark matter abundance, and under-abundant neutralinos, where the thermal relic density is smaller than the dark matter abundance. In the under-abundant case, we assume that a different particle species constitutes the remainder of the dark matter density. Indicating the thermal neutralino density with $\rho_\chi$ and the cosmological dark matter density with $\rho_{\rm DM}$, we assume that the ratio $\xi=\rho_\chi/\rho_{\rm DM}\le 1$ is constant everywhere in the universe. As a result we rescale the DM direct detection rates by the factor $\xi$, and the gamma-ray flux from neutralino pair-annihilation by a factor $\xi^2$. We assume a standard cosmological thermal history, and thus we rule out models with $\xi>1$, even though a late episode of entropy injection or other modifications to the thermal history could in principle make such over-abundant models viable.

In this paper we focus on the MSSM and explore how not-so-well tempered neutralinos are constrained by current and future direct and indirect dark matter detection experiments, as well as by collider searches for supersymmetric particles. The standard lore, which we reviewed above, is that direct detection largely rules out well-tempered bino-higgsino scenarios. Here we bring a different perspective compared to other previous studies; specifically, we\footnote{Some of these phenomenological features have been observed in global fit studies of supersymmetric models, see e.g.~Refs.~\cite{Bagnaschi:2015eha,Bechtle:2015nua,Athron:2016gor,Athron:2017qdc,Athron:2017yua}.}
\begin{itemize}
\item[(i)] include under-abundant models, where direct detection constraints are suppressed relative to collider searches by a factor $\xi$;
\item[(ii)] explore the impact of relatively light non-standard Higgs bosons. The quasi-on-shell $s$-channel exchange of the CP-odd and CP-even heavy Higgs boson opens well-known \emph{funnels} in the parameter space, where the thermal relic density can be significantly suppressed by resonant DM pair-annihilation;
\item[(iii)] systematically compare current and future DM direct detection constraints with current and projected limits from collider searches for supersymmetric particles and with current and future gamma-ray searches.
\end{itemize}

The remainder of the paper is structured as follows: in Sec.~\ref{sec:welltemperedneutralino} we review the neutralino sector of the MSSM and fix the notation for the relevant masses and mixing parameters; we then describe the bino-Higgsino and bino-wino mixed neutralino cases. Sec.~\ref{sec:constraints} describes the current constraints that we include in our study, specifically, DM direct detection experiments, DM searches with gamma rays, and collider searches for electroweakinos. Finally, we present our results in Sec.~\ref{sec:results} and conclude in Sec.~\ref{sec:conclusions}.

\section{Well-tempered neutralino scenarios}
\label{sec:welltemperedneutralino}
In this section we briefly review the neutralino sector of the MSSM and the specific cases of well-tempered bino-Higgsino and bino-wino admixtures of the lightest neutralino. We follow the conventions and notations of Ref.~\cite{Drees:2004} (see also e.g.~Refs.~\cite{Nilles:1983ge,Haber:1984rc,Martin:1997ns,Jungman:1995df,Olive:2016xmw} for introductions to the MSSM).

\subsection{The neutralino sector of the MSSM}

In the MSSM, the fermionic superpartners of the electrically neutral $\mathrm{SU}(2)_L \times \mathrm{U}(1)_Y$ gauge bosons, namely the \emph{bino}, $\tilde{B}$ (the superpartner of the $\mathrm{U}(1)_Y$ gauge boson $B$) and the \emph{wino}, $\tilde{W}^0$ (the superpartner of the neutral $\mathrm{SU}(2)_L$ gauge boson $W^0$), mix with the superpartners of the neutral components of the two MSSM Higgs doublets, the \emph{Higgsinos}, $\tilde{H}_d^0$ and $\tilde{H}_u^0$. The mixing is governed by the mass matrix
\begin{align}
M = 
\begin{pmatrix}
M_1 & 0 & - M_Z c_\beta s_W & M_Z s_\beta s_W \\
0 & M_2 & M_Z c_\beta c_W & - M_Z s_\beta c_W \\
-M_Z c_\beta s_W & M_Z c_\beta c_W & 0 & -\mu \\
M_Z s_\beta s_W & - M_Z s_\beta c_W & - \mu & 0 
\end{pmatrix}.
\end{align}
Here and throughout this paper we use the shorthand notation $c_\beta = \cos\beta$, $s_\beta = \sin\beta$, $t_\beta = \tan\beta$, $c_W = \cos\theta_W$, $s_W = \sin\theta_W$, $t_W = \tan\theta_W$ etc.. $\theta_W$ is the weak mixing angle and $M_Z$ is the mass of the $Z$ boson. In general, the gaugino and Higgsino mass parameters $M_1$, $M_2$ and $\mu$ are complex, thus allowing for new sources of CP-violating interactions beyond the SM. We restrict ourselves to the CP-conserving MSSM in this work, thus all mass parameters are real.

After diagonalization, we obtain the physical mass eigenstates, the \emph{neutralinos}, $\neut{i}$ ($i=1,\dots,4$, enumerating the mass ordered states from lightest to heaviest). The neutralino masses are then given by
\begin{align}
\mathrm{diag} (m_{\neut{1}}, m_{\neut{2}}, m_{\neut{3}}, m_{\neut{4}} ) = Z^* M Z^{-1},
\end{align}
where $Z$ is the neutralino mixing matrix. The components of the $Z$ matrix play a central role in the coupling expressions of the neutralino interactions with the neutral Higgs bosons. The relevant terms in the Lagrangian for neutralino-neutralino-scalar interactions are
\begin{align}
\mathcal{L}_{\tilde{\chi}^0 \tilde{\chi}^0 \phi} = & - \tfrac{g_2}{2} \of{ H c_\alpha - h s_\alpha } \overline{  \tilde{\chi}^0} _n \of{ P_R Q^{''}_{n\ell} + P_L Q^{'' *}_{\ell n} } \tilde{\chi}^0_\ell \nonumber\\
& + \tfrac{g_2}{2} \of{ H s_\alpha + h c_\alpha } \overline{\tilde{\chi}^0}_n \of{ P_R S^{''}_{n\ell} + P_L S^{''*}_{\ell n} } \tilde{\chi}^0_\ell \nonumber\\
& - i \tfrac{ g_2}{2} A \overline{ \tilde{\chi}^0}_n { P_R \of{ Q^{''}_{n\ell} s_\beta - S^{'' *}_{n \ell}c_\beta } } \tilde{\chi}^0_\ell \nonumber \\
& -  i \tfrac{g_2}{2} A \overline{\tilde{\chi}^0}_n { P_L \of{ S^{''*}_{\ell n} c_\beta - Q^{'' *}_{\ell n} s_\beta } }  \tilde{\chi}^0_\ell .
\end{align}
We denote by $\phi=h$ and $H$ the light and heavy neutral CP-even Higgs boson, respectively, obtained after electroweak symmetry breaking from the diagonalization of the CP-even neutral components of the Higgs fields $H_u$ and $H_d$ by a rotation with mixing angle $\alpha$. $A$ is the neutral CP-odd Higgs boson of the MSSM Higgs sector, and $g_2$ is the $\mathrm{SU}(2)_L$ gauge coupling strength. We furthermore defined
\begin{align}
Q_{n\ell}'' \equiv \tfrac{1}{2} \left[ Z_{n3} \of{Z_{\ell2}-t_W Z_{\ell1}} + Z_{\ell3} \of{ Z_{n2} - t_W Z_{n1} }  \right],\\
S_{n\ell}'' \equiv \tfrac{1}{2} \left[ Z_{n4} \of{Z_{\ell2} - t_W Z_{\ell1}}+Z_{\ell4}\of{ Z_{n2}-t_W Z_{n1} } \right].
\end{align}
Note that if the lightest neutralino, $\neut{1}$, is either purely Higgsino ($Z_{11}=Z_{12}=0$) or has no Higgsino component ($Z_{13} = Z_{14} = 0$), then $Q_{11}'' = S_{11}'' =0$. In other words, only if the lightest neutralino has both gaugino and Higgsino components, the $\neut{1} \neut{1} \phi$ (with $\phi = h,H,A$) couplings will be nonzero. These couplings play a major role both in neutralino pair-annihilation processes as well as neutralino-nucleon scattering processes that are mediated by Higgs bosons.

\subsection{Bino-Higgsino case}\label{sec:binohiggsino}

For $M_1\sim\mu \ll M_2$, a scenario we indicate as bino-Higgsino case, in the limit of large sfermion masses and outside regions of resonant Higgs-mediated annihilation, the lightest neutralino thermal relic density is set by the interplay of four species which, depending on the relative splitting between $M_1$ and $\mu$, contribute to the effective pair-annihilation cross section via co-annihilation: the three lightest neutralinos, and the lightest chargino. Analytical expressions for the thermal relic density in the limit of small mixing and for pure Higgsinos can be found, for instance, in Ref.~\cite{ArkaniHamed:2006mb}. Above the $Z$ mass and below a lightest neutralino mass $\mneut{1}\lesssim 500\gev$, the correct thermal relic density is obtained for $\mu\gtrsim M_1$, with the difference $(\mu - M_1)$ decreasing with increasing $\mneut{1}$. For $\mneut{1} > 500\gev$, we need $M_1 \gtrsim \mu$ for the correct thermal relic abundance, which eventually saturates to the pure Higgsino value for lightest neutralino masses $\sim1\tev$. At that point $M_1$ can take arbitrarily large values, and the lightest neutralino is essentially a pure Higgsino.

\subsection{Bino-Wino case}\label{sec:binowino}
The mixed bino-wino case is rather similar to the bino-Higgsino scenario, and it occurs if $M_1\lesssim M_2\ll\mu$; the main driver of the thermal relic abundance is the degree of mixing between the lightest, bino-like neutralinos and its SU(2)-charged co-annihilating partners. Such mixing is, to lowest order, controlled by a mixing angle \cite{ArkaniHamed:2006mb}
\begin{align}
\theta\equiv\frac{\sin^2\theta_W\sin^2\beta M_Z^2}{2\mu(M_2-M_1)}.
\end{align}
The role of $\mu$ is thus crucial for both the spin-independent direct detection cross section (which depends on the degree of Higgsino times gaugino fraction of the lightest neutralino) and for setting the thermal relic density, at least for lightest neutralino masses below the mass that produces a good thermal relic in the pure wino limit, around $2\tev$. Once again, above this latter threshold, the lightest neutralino is wino-like for any value of $M_1>M_2$.

\subsection{Wino-Higgsino case}\label{sec:winohiggsino}
A final third possibility is that of mixed wino-Higgsinos, occurring for $M_2\sim\mu \ll M_1$. The multiple coannihilation channels and annihilation final states including $\mathrm{SU}(2)_L$ gauge bosons highly suppress the thermal relic density for any mass within LHC reach, i.e.\ in the hundreds of GeV range, producing highly suppressed direct detection and indirect detection rates. Collider searches are the only avenue open in this case, and the exercise of comparing such searches with dark matter searches is in this case not informative. We therefore do not consider this case here.

\section{Experimental Constraints}\label{sec:constraints}

In this section we describe the relevant experimental constraints on not-so-well tempered neutralino DM scenarios in the MSSM, and how we include them in our study. We discuss first the dark matter constraints obtained from the determination of the dark matter relic abundance, direct and indirect detection experiments, and then the collider constraints from searches for supersymmetric particles, as well as from Higgs searches, at the LEP and LHC experiments.

On a technical note, we generate the MSSM particle mass and decay spectrum with the \texttt{SUSYHIT} package (version 1.5a)~\cite{Djouadi:2006bz,Djouadi:1997yw,Muhlleitner:2003vg}. DM relevant quantities such as the neutralino relic abundance, the annihilation cross section and neutralino-nucleon scattering cross section are evaluated with \texttt{MicrOMEGAs} (version 4.3.1)~\cite{Belanger:2001fz,Belanger:2004yn}.

\subsection{Dark Matter Constraints}\label{sec:dmconstraints}
\label{sec:DM}

We assume a standard cosmological thermal history, and we assume that neutralinos are only produced through thermal freeze-out. We rule out models with a thermal relic density larger than the 95\% C.L. upper limit on the cold dark matter density in the universe, according to Ref.~\cite{Ade:2015xua}. Given that $\Omega_{\rm DM}h^2=0.1188\pm0.0022$, where $\Omega_{\rm DM}=\rho_{\rm DM}/\rho_{\rm crit}$ and $h$ is the Hubble parameter in units of 100 km/s/Mpc, the upper limit is $\Omega_{\rm DM}^{\rm max}h^2=0.1232$. For models where the thermal relic density $\rho_\chi\le\rho_{\rm DM}$ we define the quantity $\xi\equiv{\rm min}\left(1,\Omega_\chi/\Omega_{\rm DM}\right)$, thus allowing for the possibility that neutralinos are only a fraction $\xi$ of the dark matter in the universe. We assume that $\xi$ is constant at all points in the universe and consider a local neutralino density $\rho_\chi(\vec x)=\xi\rho_{\rm DM}(\vec x)$. We will consider models where $\Omega_\chi/\Omega_{\rm DM}>1$ as ruled out, even though we will show results for those models, entertaining the possibility that the right relic density is achieved via appropriate dilution of the excessively large thermal relic density. 

In practice, our procedure yields a suppression by a factor $\xi$ for direct detection rates, which depend linearly on $\rho_\chi=\xi\rho_{\rm DM}$, and a suppression by a factor $\xi^2$ for indirect detection rates, specifically gamma-ray from dark matter annihilation, which depend on $\rho_\chi^2=\xi^2\rho^2_{\rm DM}$. 

In terms of the specifics of the constraints, for the direct detection constraints we use the results of the 2017 XENON1T collaboration  \cite{Aprile:2017iyp}, which are marginally but visibly stronger, on our plots, than the recent 2016 LUX analysis \cite{Akerib:2016vxi}. We make the same assumptions on the local dark matter density, velocity distribution, quark content of the proton as in Ref.~\cite{Akerib:2016vxi, Aprile:2017iyp}.

For indirect detection with gamma rays, we consider a proxy to the photon flux that only captures the ``particle physics'' factor,
\begin{align}
\tilde{\phi}_\gamma = c_\gamma \frac{\langle \sigma v \rangle_0}{m_\chi^2},
\end{align}
where
\begin{align}
c_\gamma = \int_{E_\text{th}}^{m_\chi} \frac{dN}{dE} dE \simeq c\cdot m_\chi^\alpha
\end{align}
indicates the gamma-ray multiplicity per dark matter annihilation
with $\alpha\simeq 1.0$ and $c\simeq1.0$. We thus have
\begin{align}
\tilde\phi_\gamma \simeq\frac{\langle \sigma v \rangle_0}{m_\chi^2} m_\chi^\alpha \xi^2.
\end{align}
Following the results of Ref.~\cite{Ackermann:2015zua} on searches for gamma rays from local dwarf spheroidal galaxies, we approximate current constraints from gamma-ray observations using 
\begin{align}\label{eq:grlimits}
\phi_\gamma \lesssim \left(\frac{\langle \sigma v \rangle_0}{3 \cdot 10^{-26}\ {\rm cm}^3{\rm s}^{-1}}\right) \cdot \left(\frac{100\gev}{m_\chi}\right)^{2-\alpha}
\end{align}
Although approximate and somewhat model-dependent, this criterion mirrors the general ability of indirect searches with gamma rays to set firm constraints on the parameter space of interest here. Other indirect or astrophysical searches for dark matter are generally more model-dependent, or less sensitive, than the limits implied by Eq.~\eqref{eq:grlimits}.

\subsection{Collider Constraints}\label{sec:colliders}

\subsubsection{\LEP\ Searches for EW Gauginos}
\label{sec:LEP}
The combination of results from all four LEP experiments from searches for chargino pair production yields a lower limit on the chargino mass of $\mcharg{1}\ge 103.5\gev$~\cite{LEPcombinationchargino}.
This limit is quite robust, and only weakens slightly, to $\mcharg{1} \gtrsim 95\gev$, if the mass splitting $\Delta m$ between the chargino and the lightest neutralino is below $3\gev$ (reaching the weakest limit of $\mcharg{1} \gtrsim 92\gev$ at $\Delta m \approx 200\MeV$)~\cite{LEPcombinationchargino_smalldm}.
However, in the results shown in this work $\Delta m$ always exceeds $3\gev$, so that we can apply the limit $\mcharg{1}\ge 103.5\gev$ everywhere.\footnote{Scenarios with $\Delta m < 3\gev$ and $\mcharg{1}\sim \order{90-100}\gev$ can only be obtained if the bino component of the LSP is very small, and thus generically lead to a neutralino relic density that is much too small compared to the observed dark matter density.}

\subsubsection{\LHC\ Searches for EW Gauginos}
\label{sec:LHCsusy}

CMS has searched for electroweak gauginos with two soft oppositely charged leptons (electrons, muons), missing transverse energy and an initial state radiated (ISR) jet in the final state~\cite{CMS:2016zvj,CMS:2017fij}, following ideas from Refs.~\cite{Schwaller:2013baa,Han:2014kaa} (see also Ref.~\cite{Gori:2013ala}). This search is particularly sensitive to scenarios with small mass splittings $\Delta m$ down to $\sim 7.5\gev$ between the decaying EW gaugino and the lightest neutralino. A simplified model interpretation is given for the production of the lightest chargino, $\tilde{\chi}_1^\pm$, and second lightest neutralino, $\tilde{\chi}_2^0$, each assumed to be pure wino gauge eigenstates, and fully decaying to the lightest neutralino, $\neut{1}$, and a $W^\pm$ or $Z$ boson, respectively. 

The same SUSY process and simplified model was also searched for in multilepton final states by ATLAS~\cite{ATLAS:2016uwq} and CMS~\cite{CMS:2016gvu,CMS:2017fdz}. These searches become only sensitive for mass splittings of roughly $\Delta m \gtrsim 25\% \cdot \mcharg{1}$ due to the minimal $p_T$ requirements of the leptons of the order of $20$ to $25\gev$. They exhibit their full sensitivity for large mass splittings $\Delta m$ well above the $Z$ boson mass, $M_Z$.

In our analysis we include the CMS upper cross section limits based on the latest results from the Moriond 2017 conference, using $35.9\invfb$ of $13\tev$ data~\cite{CMS:2017fij,CMS:2017fdz}. We evaluate the $\chargpm{1}\neut{2}$ (and, in the bino-Higgsino scenario, $\chargpm{1}\neut{3}$) production cross section at next-to-leading order (NLO) including the resummation of large logarithms at the next-to-leading log (NLL) level with the public code \texttt{Resummino-2.0.1}~\cite{Debove:2009ia,Debove:2010kf,Debove:2011xj,Fuks:2013vua,Fuks:2012qx}, using \texttt{CTEQ6.6LO}~\cite{Nadolsky:2008zw} and \texttt{MSTW2008NLO90CL}~\cite{Martin:2009iq} parton distribution functions (PDFs) through the \texttt{LHAPDF} framework~\cite{Buckley:2014ana}. 

We furthermore include approximations of the expected limits for future integrated luminosities of $100\invfb$ and $300\invfb$ in our analysis, assuming future null results and that the total uncertainty is statistically dominated.
Note, however, that the accuracy of our future limit estimates is rather limited due to the lack of public information from the CMS search analyses.\footnote{In particular, the expected upper cross section limit, $\sigma_\text{UL}^\text{exp}$, is not publicly available in the $\chargpm{1}$-$\neut{1}$ mass plane. We therefore estimated $\sigma_\text{UL}^\text{exp}$ by rescaling the observed upper cross section limit, $\sigma_\text{UL}^\text{obs}$, by the ratio $\sigma/\sigma_\text{UL}^\text{obs}$ calculated at the $\CL{95}$ expected limit contour line.}

\subsubsection{LHC Searches for Higgs bosons}
\label{sec:LHCHiggs}

In the scenarios we consider here, we assume that SUSY and Higgs mixing effects are sufficiently decoupled such that the light Higgs boson possesses SM Higgs couplings to very good approximation. Hence, the LHC Higgs boson signal strength measurements cannot discriminate our scenarios from the SM picture unless the light Higgs boson can decay into light SUSY states, in particular, to two lightest neutralinos, $h\to \neut{1}\neut{1}$, which is an invisible final state. Taking the combined ATLAS and CMS Higgs boson signal strength determination for the $125\gev$ Higgs boson~\cite{Khachatryan:2016vau}, $\hat\mu = 1.09 \pm 0.11$, we can infer a $\CL{95}$  upper limit on the invisible branching fraction of the Higgs boson, $\mathrm{BR}(h\to \neut{1}\neut{1}) \le 14.4\%$, assuming the Higgs couplings are unaltered with respect to the SM (see e.g.~Ref.~\cite{Belanger:2013kya,Bechtle:2014ewa} for the methodology).
Direct LHC searches for invisible Higgs decays, assuming SM Higgs production rates, currently give weaker upper limits on $\mathrm{BR}(h\to \neut{1}\neut{1})$, namely $28\%$ and $36\%$ from ATLAS~\cite{Aad:2015txa} and CMS~\cite{CMS:2015naa}, respectively.
It turns out that in all scenarios investigated in this work, $\mathrm{BR}(h\to \neut{1}\neut{1})$ never exceeds the experimentally allowed value.

Since we focus in particular on scenarios where the heavier MSSM Higgs bosons $H$ and (to a lesser extent) $A$ in the intermediate mass range\footnote{We choose $M_H \sim M_A = 500\gev$ and $\tan\beta=10$ as illustrative example values in Sec.~\ref{sec:results}.} influence the neutralino DM phenomenology, we need to comment on the constraints arising from LHC searches for these Higgs bosons. At large values of $\tan\beta$ the strongest constraints arise from searches for $pp\to H/A \to \tau^+\tau^-$~\cite{ATLAS:2016fpj,CMS:2016rjp}, where the Higgs bosons $H$ and $A$ are produced either through gluon fusion or in association with a pair of bottom quarks. The $H$ and $A$ couplings to down-type fermions are $\tan\beta$-enhanced, leading to sizable signal rates for this process. For instance, in the representative MSSM $m_h^\text{mod+}$ benchmark scenario~\cite{Carena:2013ytb}, current null results from ATLAS~\cite{ATLAS:2016fpj} and CMS~\cite{CMS:2016rjp} set an upper limit of $\tan\beta\lesssim 15$ at $M_A = 500\gev$. Note, however, that in this scenario the gaugino and Higgsino soft-breaking mass parameters are fixed to $2 M_1 \approx M_2 = \mu = 200\gev$, leading to a light electroweakino mass spectrum and thus to non-vanishing heavy Higgs to light neutralino/chargino decays. Consequently, this suppresses the branching fraction for the $H$ and $A$ decays into $\tau^+\tau^-$, leading to a weaker $\tan\beta$ limit than in scenarios where Higgs decays to electroweakinos are absent. In our scenarios the parameters $M_1$, $M_2$ and $\mu$ vary, and we thus employ the code \texttt{HiggsBounds-5.1.0beta}~\cite{Bechtle:2008jh,Bechtle:2011sb,Bechtle:2013gu,Bechtle:2013wla,Bechtle:2015pma} to check the constraints from these (and other) Higgs searches in the full parameter space.

\begin{figure*}[ht!]
\centering
\includegraphics[width=0.5\textwidth]{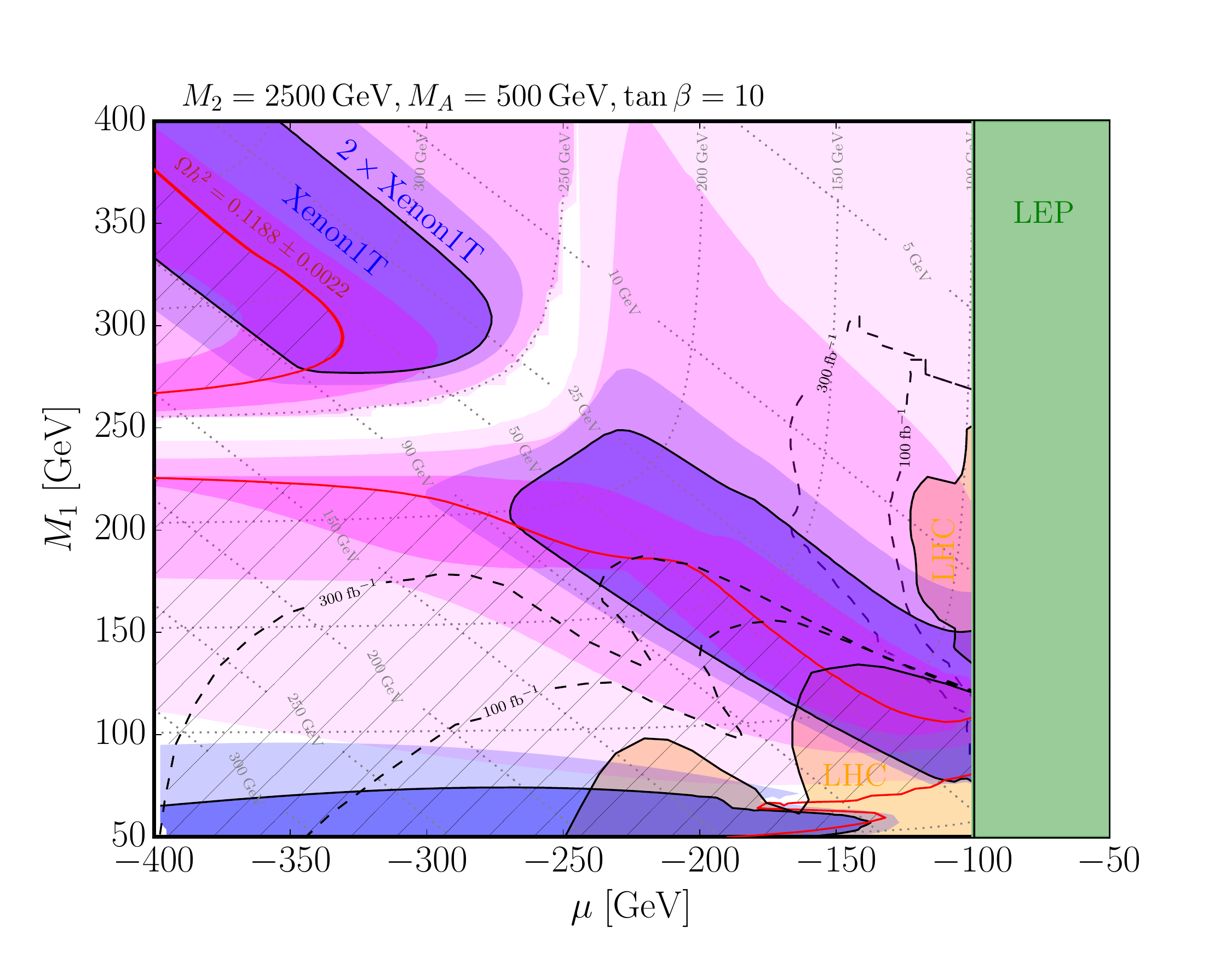}\hfill
\includegraphics[width=0.5\textwidth]{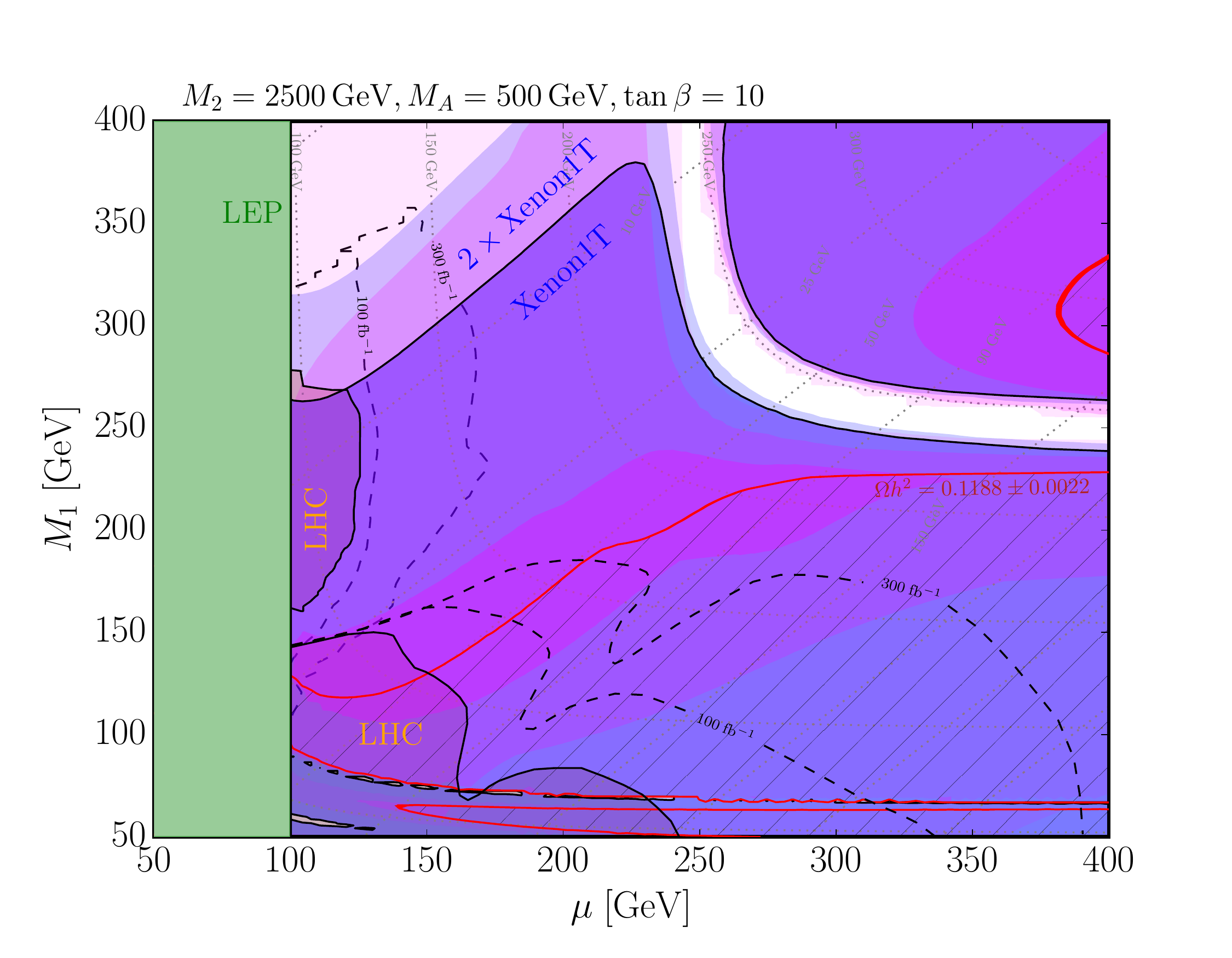}
\caption{Comparison of current and projected constraints in the well-tempered bino-Higgsino neutralino DM scenario in the ($\mu$, $M_1$) parameter plane for negative $\mu$ (\emph{left}) and positive $\mu$ (\emph{right}), with $\tan\beta = 10$, $M_2 = 2.5\tev$ and a pseudoscalar Higgs mass of $M_A = 500\gev$. The orange regions are excluded by two different LHC searches for $pp\to \chargpm{1}\neut{2} \to W^\pm Z\neut{1}\neut{1}$ by the CMS experiment~\cite{CMS:2017fij,CMS:2017fdz} (see text). Projected limits for $100\ifb$ and $300\ifb$ of the CMS searches are shown by dashed lines. The exclusion obtained from the \LEP\ chargino mass limit is shown in green. Current and projected (improvement of a factor 2) exclusion limits by the DM direct detection experiment \XenonT~\cite{Aprile:2017iyp} are shown as blue shaded areas with varying opacity (see blue text labels). Projected limits from DM indirect detection experiments are indicated by the magenta regions, where the current limit is scaled by factors of 10, 100 and 1000, with high to low opacity. The narrow region that predicts the observed DM relic density is shown in red. An overlaid hatching marks the over-abundant neutralino DM regions, assuming a standard cosmological thermal history.
The roughly diagonal and hyperbolic-like gray dotted contours give values for the mass splitting, $\Delta m = \mcharg{1} - \mneut{1}$, and the DM mass, $\mneut{1}$, respectively.}\label{fig:mum1}
\end{figure*}

Our example choice of $M_A=500\gev$ and $\tan\beta=10$ is still allowed by current limits from LHC $pp\to H/A \to \tau^+\tau^-$ searches. The parts of the parameter space where Higgs decays to electroweakino are kinematically forbidden will however be excluded if the current limit is improved by $\sim 25\%$, whereas in the experimentally allowed parameter regions with maximal branching fractions of the heavier Higgs bosons to electroweakinos (see Sec.~\ref{sec:results_binoHiggsino}) an improvement by $\sim (50-60)\%$ is needed to obtain exclusion. Yet, as we will argue at the end of Sec.~\ref{sec:results}, the dark matter phenomenology and Higgs funnel mechanism is only mildly dependent on $\tan\beta$, thus the scenarios discussed in this work can easily be obtained at lower $\tan\beta$ values, which puts the $pp\to H/A \to \tau^+\tau^-$ searches out of reach in the near future.

We want to close this discussion with a comment on the possibility of very light neutralino dark matter with $\mneut{1}\lesssim  150\gev$ that achieves the correct relic density dominantly through the  Higgs funnel mechanism. In that case, the pseudoscalar Higgs mass needs to be $M_A \simeq 2\mneut{1} \lesssim 300\gev$, such that the heavy Higgs states are non-decoupled and Higgs mixing effects will impact  the light Higgs signal rates and coupling properties. In the bino-Higgsino case, where $\mu$ is small, Higgs rate measurements will significantly constrain these scenarios. In contrast, in the bino-wino case, where $\mu$ is large, the possibility of Higgs alignment without decoupling~\cite{Gunion:2002zf,Craig:2013hca,Carena:2013ooa,Carena:2014nza,Bechtle:2016kui,Haber:2017erd} arises, in which one of the CP-even Higgs bosons obtains the coupling properties of the SM Higgs boson due to an accidental cancellation of tree-level and higher-order corrections in the MSSM Higgs sector. In that case, even very light neutralino dark matter in the sub-GeV to $65\gev$ range could be viable in a scenario where the heavy CP-even Higgs boson is the SM-like Higgs boson at $125\gev$~\cite{Profumo:2016zxo}.

\section{Results}
\label{sec:results}

\subsection{Bino-Higgsino case}
\label{sec:results_binoHiggsino}
We start our study of the not-so-well tempered neutralino with the case of a bino-Higgsino admixture. We decouple the wino soft-breaking mass parameter, $M_2$, by setting it to $2.5\tev$. In order to illustrate the effects of resonant, quasi-on-shell exchange of a heavy Higgs boson we set the pseudoscalar Higgs mass to $M_A=500$ GeV and  $\tan\beta=10$. We furthermore decouple the sfermion and gluino mass parameters by setting them to $5\tev$. We choose the trilinear soft-breaking parameters (in particular $A_t$) such that a light Higgs boson mass of $M_h \simeq 125\gev$ is ensured to a good approximation everywhere in the parameter space.

We present our results in Fig.~\ref{fig:mum1}. The left panel shows our results for negative relative sign between $M_1$ and $\mu$, while the right panel displays them for positive relative sign. The phenomenology is strikingly different in the two cases due to important interference effects, as we discuss below.

The green vertical bands indicate the region ruled out by \LEP\ searches for chargino pair production (see~Sec.~\ref{sec:LEP}). The orange area is ruled out by current \LHC\ searches for $pp\to \chargpm{1}\neut{2} \to W^\pm Z\neut{1}\neut{1}$ conducted by the \CMS\ collaboration (see Sec.~\ref{sec:LHCsusy}). Here, in the parameter region at $M_1\sim (130-250)\gev$, $ |\mu| \sim (100-125)\gev$, the exclusion is obtained from the search designed for compressed electroweakino mass spectra~\cite{CMS:2017fij}, while the other regions with $M_1 \lesssim |\mu|$ are excluded by the (traditional) multi-lepton plus $\etmiss$ search~\cite{CMS:2017fdz}. Estimates of future exclusions by these collider searches with $100\ifb$ and $300\ifb$ are shown as dashed black lines.
Dark blue regions are ruled out by the recent \XenonT\ results~\cite{Aprile:2017iyp} and the light blue regions indicate the projected exclusion assuming a factor 2 improvement of the \XenonT\ limit. Current gamma-ray indirect detection experiments do not provide any constraint on this parameter space. The magenta/pink shaded regions display a projected exclusion assuming an improvement by one, two and three orders of magnitude of the current \FermiLAT\ limit obtained from dwarf-spheroidal galaxies~\cite{Ackermann:2015zua}, (see Eq.~\eqref{eq:grlimits} in Sec.~\ref{sec:DM}). The narrow region that predicts the observed DM relic density is shown in red.
For better orientation, we also include in Fig.~\ref{fig:mum1} dotted gray contour lines for the mass difference between the lightest chargino and the lightest neutralino, $\Delta m = \mcharg{1} - \mneut{1}$ (roughly diagonal lines) and of the lightest neutralino mass (hyperbolic-like lines).

The plots illustrate that without resonant annihilation, i.e.\ outside the Higgs funnel region, models with the correct thermal relic density are firmly excluded by direct detection, at least in the region where the electroweakinos are relatively light and possibly accessible at the LHC. In fact, we find that for positive relative sign between $M_1$ and $\mu$  even the funnel scenario with correct thermal relic abundance is ruled out by direct detection constraints, at least for the particular choice of $M_A$ we made here, and unless $\mu$ is extremely large (far beyond the range shown in Fig.~\ref{fig:mum1}).

Even after rescaling the direct detection exclusion limits with the thermal relic abundance of the neutralino (if $\xi < 1$), in case of positive relative sign between $M_1$ and $\mu$, current \LHC\ searches do not provide a significant additional constraint given the \XenonT\ limit. The picture is quite different for  negative relative sign between $M_1$ and $\mu$, where the spin-independent DM direct detection rates are suppressed by destructive interference between the light and heavy CP-even Higgs exchange (\emph{blind-spot}). Here, current LHC searches exclude regions not otherwise excluded by direct detection. Specifically, the \CMS\ search for compressed electroweakino mass spectra~\cite{CMS:2017fij} excludes the neutralino DM under-abundant region where the lightest neutralino is light, $\mneut{1}\sim (100-120)\gev$, and the mass splitting to the lightest chargino is relatively small, $\Delta m \sim (7.5 - 20)\gev$. We estimate that the limit will marginally improve with $100\ifb$ and will eventually reach scenarios with lightest neutralino masses above $150\gev$ with $300\ifb$.\footnote{The projected exclusion line for $100\ifb$ is only marginally better than the current exclusion due to an under-fluctuation of the currently observed signal yield (and thus a better observed limit than expected limit)~\cite{CMS:2017fij}.} The multilepton plus $\etmiss$ search~\cite{CMS:2017fdz} excludes the parameter space mostly in regions where the neutralino DM is over-abundant --- assuming the standard cosmological thermal history --- except for a region $\mu \sim -(100-150)\gev$, $M_1 \sim (110-125)\gev$ (which is however also excluded by \XenonT) and a region at low $M_1$ values, where $\mneut{1} \lesssim M_h/2 \simeq 62.5\gev$ and thus the light Higgs funnel mechanism is effective.

Prospects for indirect detection under the present assumptions are generally very bleak especially because of the suppression of rates with the inverse square of the (under-abundant) relic density. Nevertheless, in the case of negative relative sign between $M_1$ and $\mu$, and given the occurrence of a \emph{blind-spot} cancellation as chosen here, indirect detection remains the only tool to experimentally probe the heavy Higgs funnel region at 
$|\mu| > M_1$. Our conclusions would of course be different assuming non-thermal production of neutralinos in addition to the thermal relic population.

It should be noted that the \XenonT\ exclusion in Fig.~\ref{fig:mum1}(left) in the mostly over-abundant region, $M_1 < |\mu|$, depends on the details of the destructive interference between the $h$- and $H$-mediated diagrams for spin-independent DM-nucleon scattering, and thus depends on $\tan\beta$ and the heavy Higgs masses, $M_H\sim M_A$ (see e.g. Ref.~\cite{Han:2013gba}). In contrast, the LHC constraints are very robust in this regard and depend only marginally on $\tan\beta$ through its effect on the neutralino and chargino masses and mixing. 

We summarize that a sub-TeV bino-Higgsino neutralino DM candidate must be highly under-abundant unless moderately light non-standard Higgs bosons $H$ and $A$ exist and serve as quasi-resonant mediator(s) for neutralino pair-annihilation \emph{and} provide a blind-spot cancellation in spin-independent direct detection experiments. LHC searches for direct production of electroweakinos give additional constraints (besides direct detection and DM relic density) only for a very light neutralino LSP in the under-abundant neutralino DM region. 
Given these findings, we now investigate complementary LHC strategies that could further shed light on the question whether such non-standard Higgs bosons indeed exist and interact in the described way with the electroweakinos.
In particular, we assess the size of the branching ratios of non-standard Higgs boson decays into lighter electroweakino states for our scenarios. More detailed phenomenological work on these signatures can be found e.g.\ in Refs.~\cite{Gunion:1987ki,Gunion:1988yc,Djouadi:1996mj,Belanger:2000tg,Bisset:2000ud,Bisset:2007mi,Arhrib:2011rp,Li:2013nma,Belanger:2015vwa,Djouadi:2015jea,Heinemeyer:2015pfa,Barman:2016kgt}. There are also early experimental studies of the LHC discovery reach by ATLAS~\cite{ATLAS:2009zmv} and CMS~\cite{Moortgat:2001pp}. 

\begin{figure*}[t!]
\centering
\includegraphics[width=0.5\textwidth]{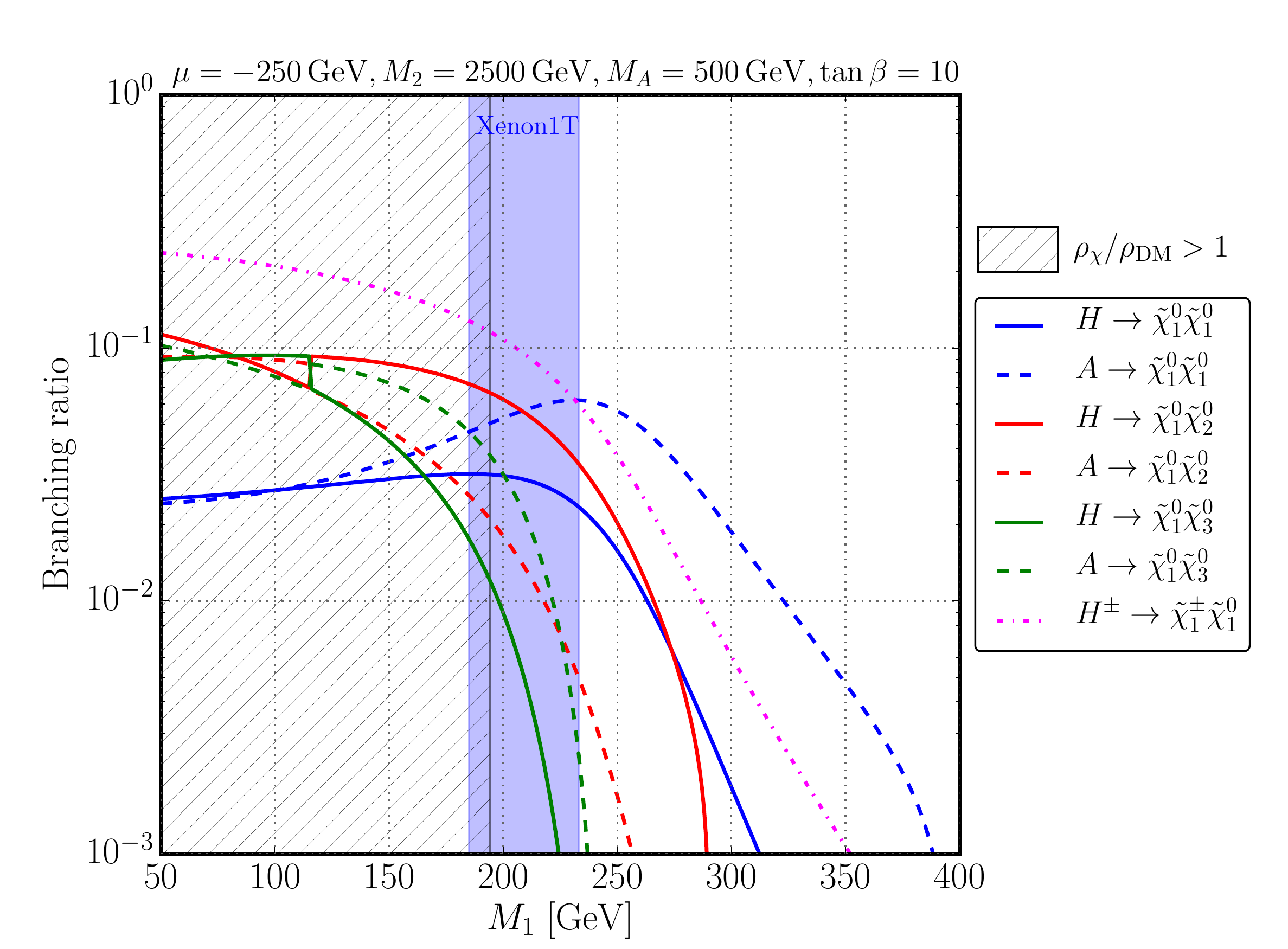}\hfill
\includegraphics[width=0.5\textwidth]{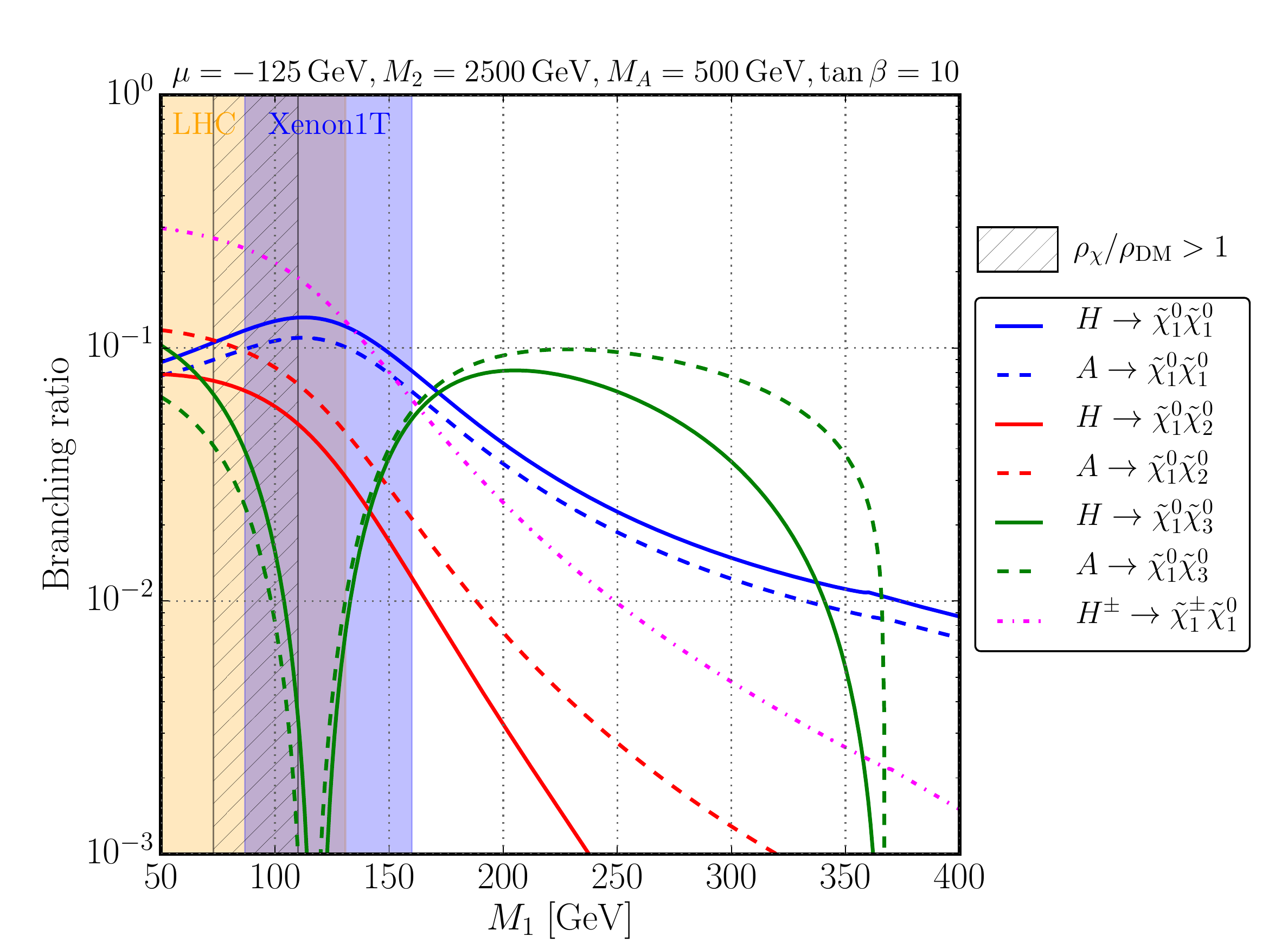}
\includegraphics[width=0.5\textwidth]{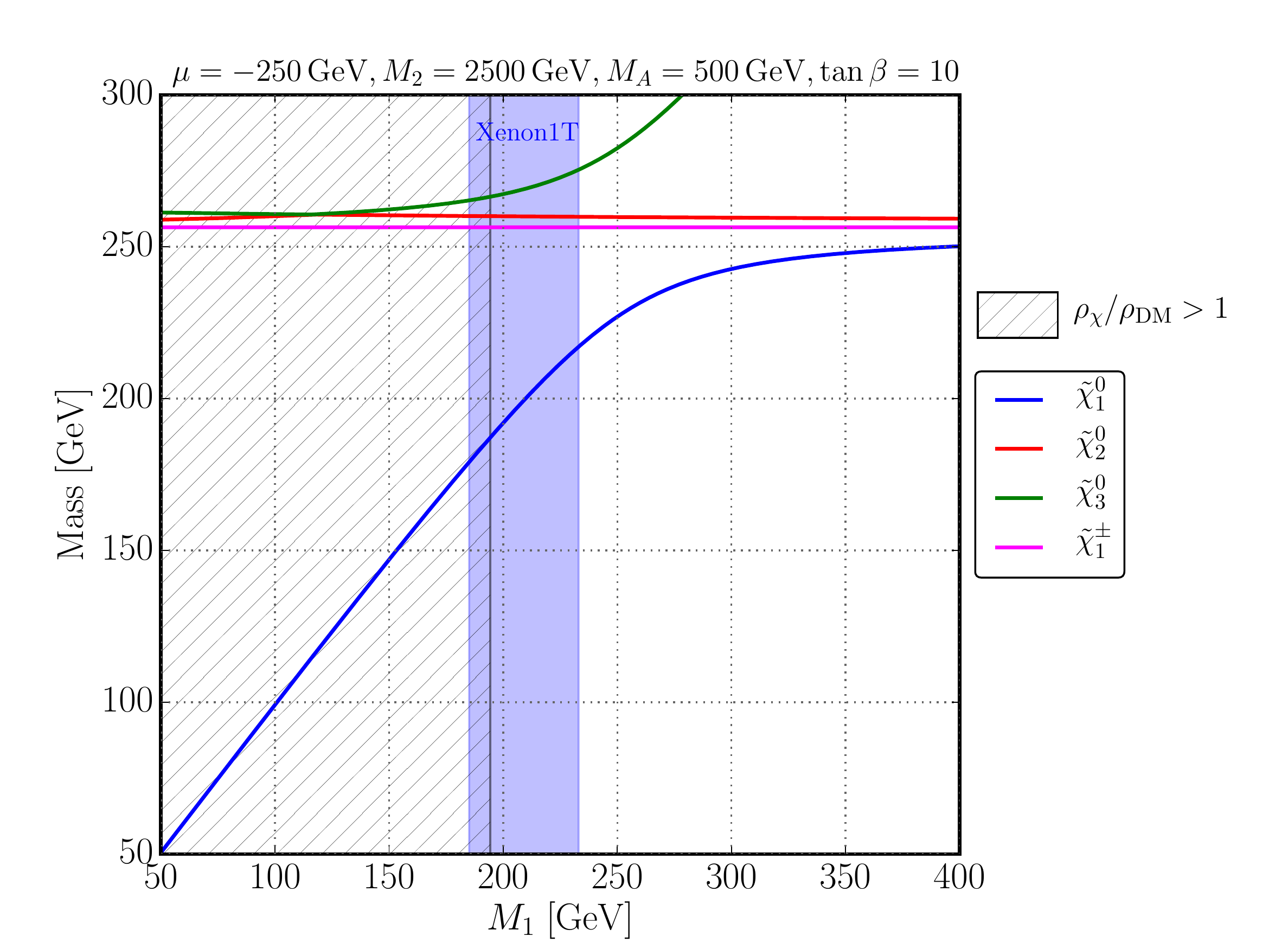}\hfill
\includegraphics[width=0.5\textwidth]{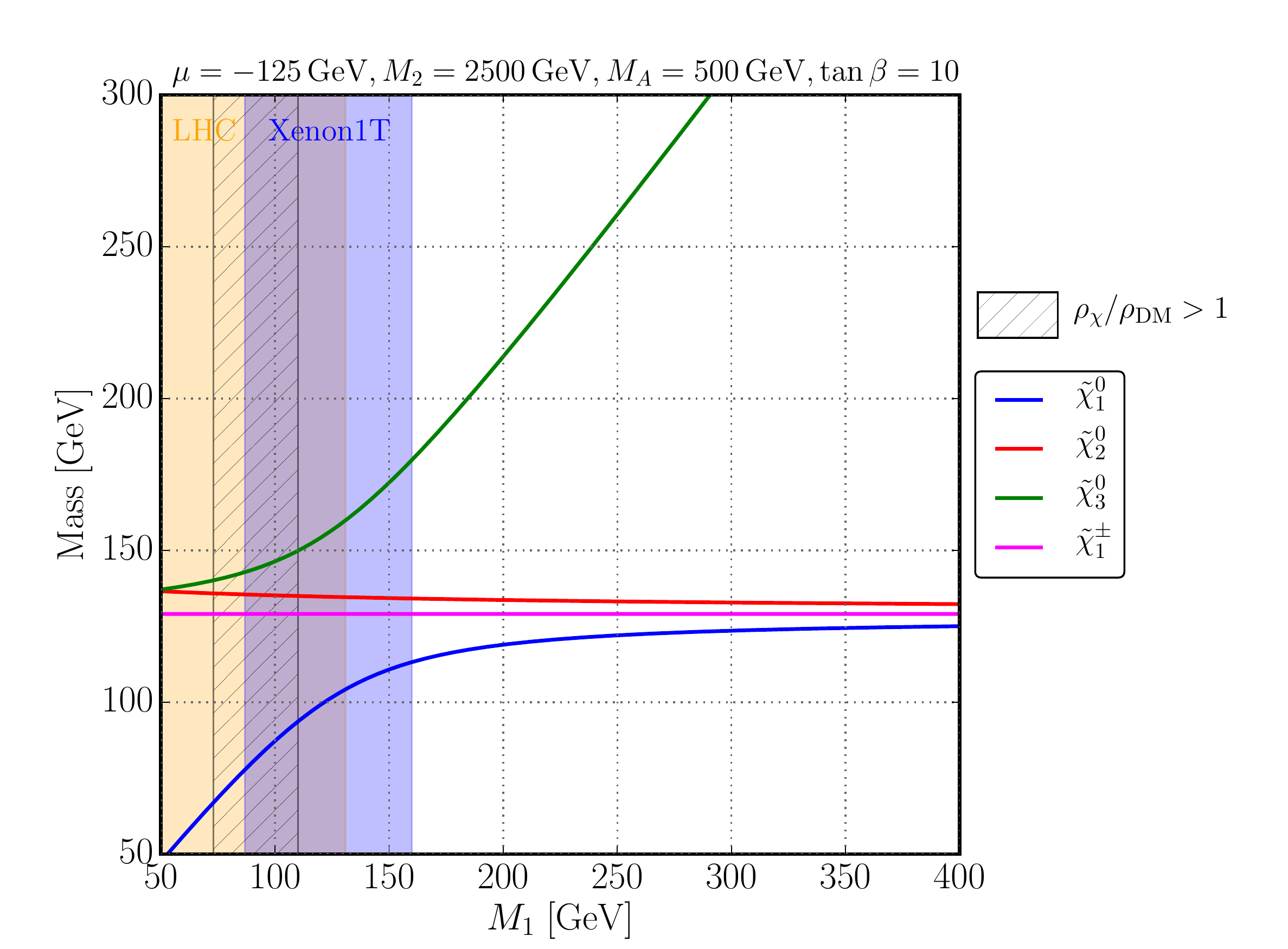}
\caption{Branching ratios of heavy Higgs boson decays to light electroweakinos (\emph{top row}) and electroweakino mass spectrum (\emph{bottom row}) as function of the bino mass parameter $M_1$ in the bino-Higgsino neutralino DM scenario, for  $\mu = -250\gev$ (\emph{left}) and $\mu = -125\gev$ (\emph{right}). We again choose $\tan\beta = 10$ and a pseudoscalar Higgs mass of $M_A = 500\gev$.}
\label{fig:binoHiggsino_Higgspheno1}
\end{figure*}

\begin{figure}[t]
\centering
\includegraphics[width=0.5\textwidth]{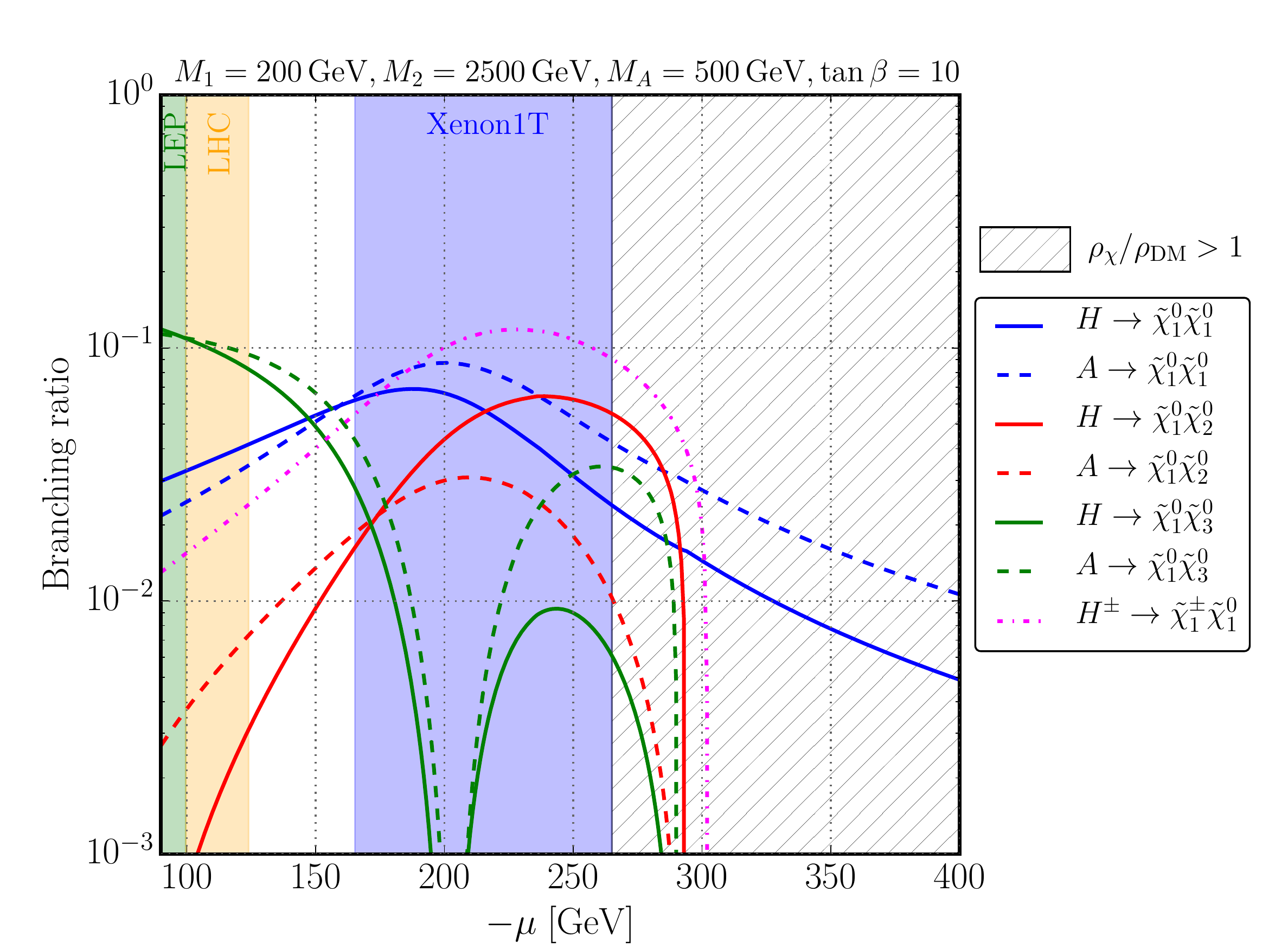}\hfill
\includegraphics[width=0.5\textwidth]{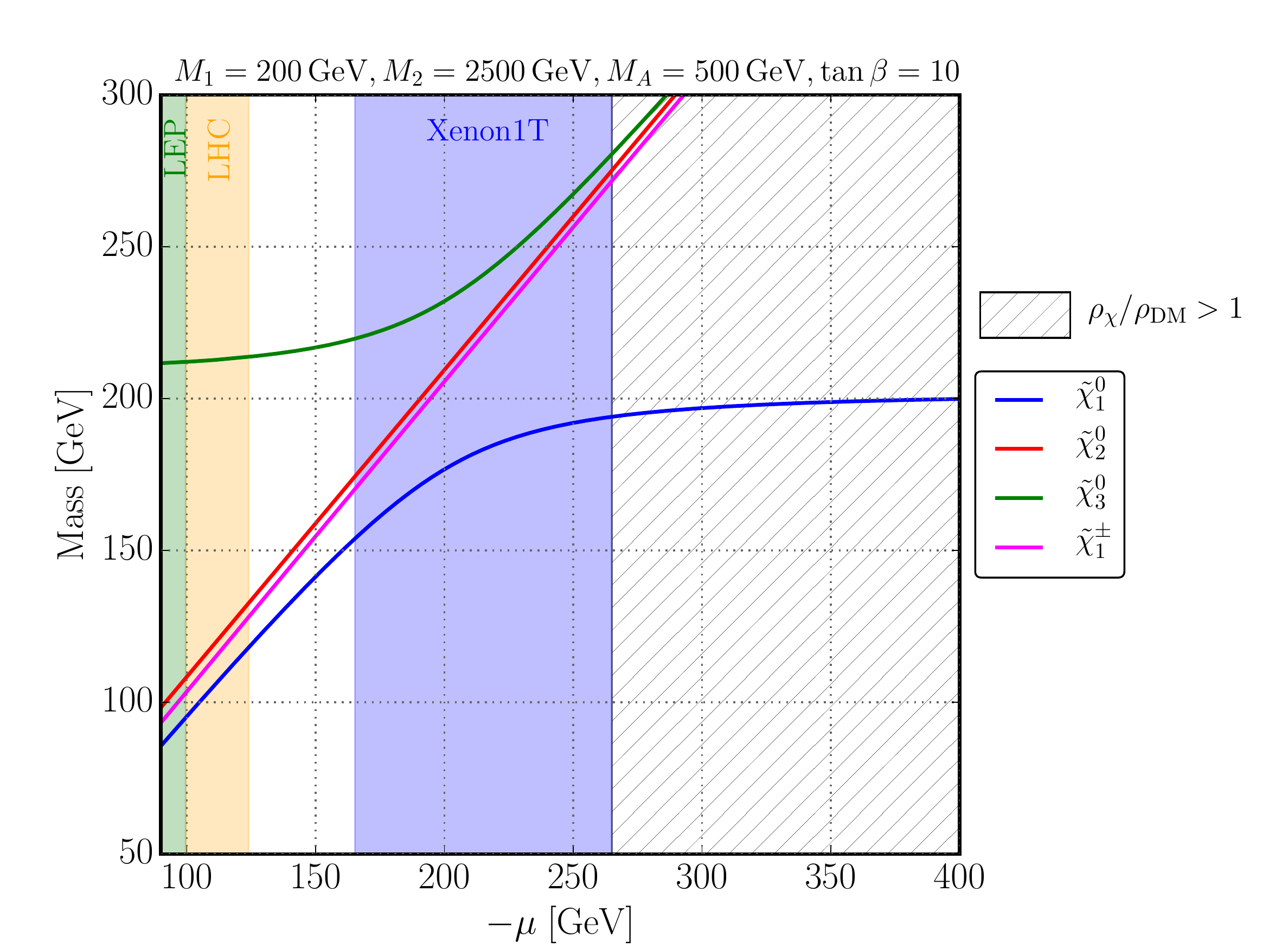}
\caption{Branching ratios of heavy Higgs boson decays to light electroweakinos (\emph{top row}) and electroweakino mass spectrum (\emph{bottom row}) as function of the Higgsino mass parameter $\mu$ in the bino-Higgsino neutralino DM scenario, for  $M_1 = 200\gev$. We again choose $\tan\beta = 10$ and a pseudoscalar Higgs mass of $M_A = 500\gev$.}
\label{fig:binoHiggsino_Higgspheno2}
\end{figure}

In the top panels of Figs.~\ref{fig:binoHiggsino_Higgspheno1} and~\ref{fig:binoHiggsino_Higgspheno2} we show the branching ratios for $H$, $A$ and $H^\pm$ decays into the lightest neutralino ($\neut{1}$) and another, heavier electroweakino ($\neut{2}$, $\neut{3}$, $\chargpm{1}$), for three different slopes through the parameter space of Fig.~\ref{fig:mum1}(left), i.e.\ for negative relative sign between $M_1$ and $\mu$. In Fig.~\ref{fig:binoHiggsino_Higgspheno1}(left) and (right) we set $\mu=-250\gev$ and $-125\gev$, respectively, and show the results as function of $M_1$. In contrast, in Fig.~\ref{fig:binoHiggsino_Higgspheno2} we fixed $M_1 = 200\gev$ and leave $\mu$ as a free parameter. The remaining parameters are the same as in Fig.~\ref{fig:mum1}. In each case, the bottom panels show the masses of the light electroweakinos ($\neut{1}$, $\neut{2}$, $\neut{3}$, $\chargpm{1}$) for the same parameter choices as in the panel above. The blue and orange shadings again indicate the excluded parameter space by \XenonT\ and \CMS, respectively, and the hatched region is excluded by the thermal relic abundance exceeding the observed dark matter abundance.

Before imposing the direct detection and relic density constraints, we find that for our scenarios (with $\tan\beta=10$) the branching fractions for the neutral Higgs decays $H/A \to \neut{1}\neut{i}$ ($i=1,2,3$) each can at most be $\sim10\%$, while the one for the charged Higgs decay $H^+ \to \neut{1}\chargp{1}$ can be slightly higher, up to $25\%$. However, after imposing the constraints from the thermal relic abundance and \XenonT\ the latter decay rate shrinks to $\sim 6\%$. The decay rate for $H/A \to \neut{1}\neut{2}$ is always significantly smaller than the rates for the invisible decay $H/A\to \neut{1}\neut{1}$ and $H/A\to \neut{1}\neut{3}$ (if kinematically allowed). The vanishing branching ratio for $H/A \to \neut{1}\neut{3}$ at $M_1 \sim 120\gev$ in Fig.~\ref{fig:binoHiggsino_Higgspheno1} is due to an accidental cancellation in the coupling expression for the $\phi \neut{1}\neut{3}$ ($\phi = H,A$) coupling.

The experimental signatures arising from these decays are the following:
The process $pp \to H/A \to \neut{3}\neut{1}$, with successive decay $\neut{3} \to Z \neut{1}$, offers the opportunity to search for a $Z$+$\etmiss$ signature. In the yet unexcluded regions (see above), the mass splitting between $\neut{3}$ and $\neut{1}$ is at least $\sim 50\gev$, thus leptons originating from the decaying $Z$ boson should be hard enough to be experimentally observable. Alternatively, the decay $\neut{3}\to h \neut{1}$ can give rise to a $h + \etmiss$ signature. We found that branching ratios of up to $60\%$ are possible in some regions of the parameter space [e.g.~at around $M_1 \sim (110-115)\gev$ for $\mu=-250\gev$ (left panels in Fig.~\ref{fig:binoHiggsino_Higgspheno1})]. For $\tan\beta \gtrsim 10$, the Higgs production in association with $b$-quarks is also sizable, leading to the possibility of additional $b$-jets. In our example for $M_A = 500\gev$, $\tan\beta = 10$, we estimate the $13\tev$ LHC cross section for $H$ ($A$) production to be around $\sim 34.8~(45.5)\fb$ and $\sim 229.9~(230.6)\fb$ for gluon fusion and bottom quark associated Higgs production, respectively.\footnote{Our cross section estimates are obtained by rescaling the most recent predictions from the LHC Higgs Cross Section Working Group, see Ref.~\cite{deFlorian:2016spz} (and references therein).} In addition, the decay $H/A \to \neut{2}\neut{1}$ leads to the same signature and will contribute to the signal yield, however, the branching fraction is smaller and the $\neut{2}$--$\neut{1}$ mass splitting is smaller. Searching for the invisible Higgs decay $H/A \to \neut{1}\neut{1}$ is a rather difficult as this signature requires at least one additional object, such as a $Z$ boson from the Higgs-Strahlungsprocess, a jet from initial state radiation, or $b$-jets from Higgs-bottom quark associated production. Lastly, experimental searches for the charged Higgs boson decay $H^\pm \to \chargpm{1}\neut{1}$, with the chargino decaying to $100\%$ to a $W$ boson and the neutralino LSP, can be performed assuming top-quark associated charged Higgs production, $pp \to H^\pm tb$, although SM backgrounds from top pair production are challenging~\cite{Bisset:2000ud}.

\subsection{Bino-Wino case}

\begin{figure*}[!h]
\centering
\mbox{\includegraphics[width=0.5\textwidth]{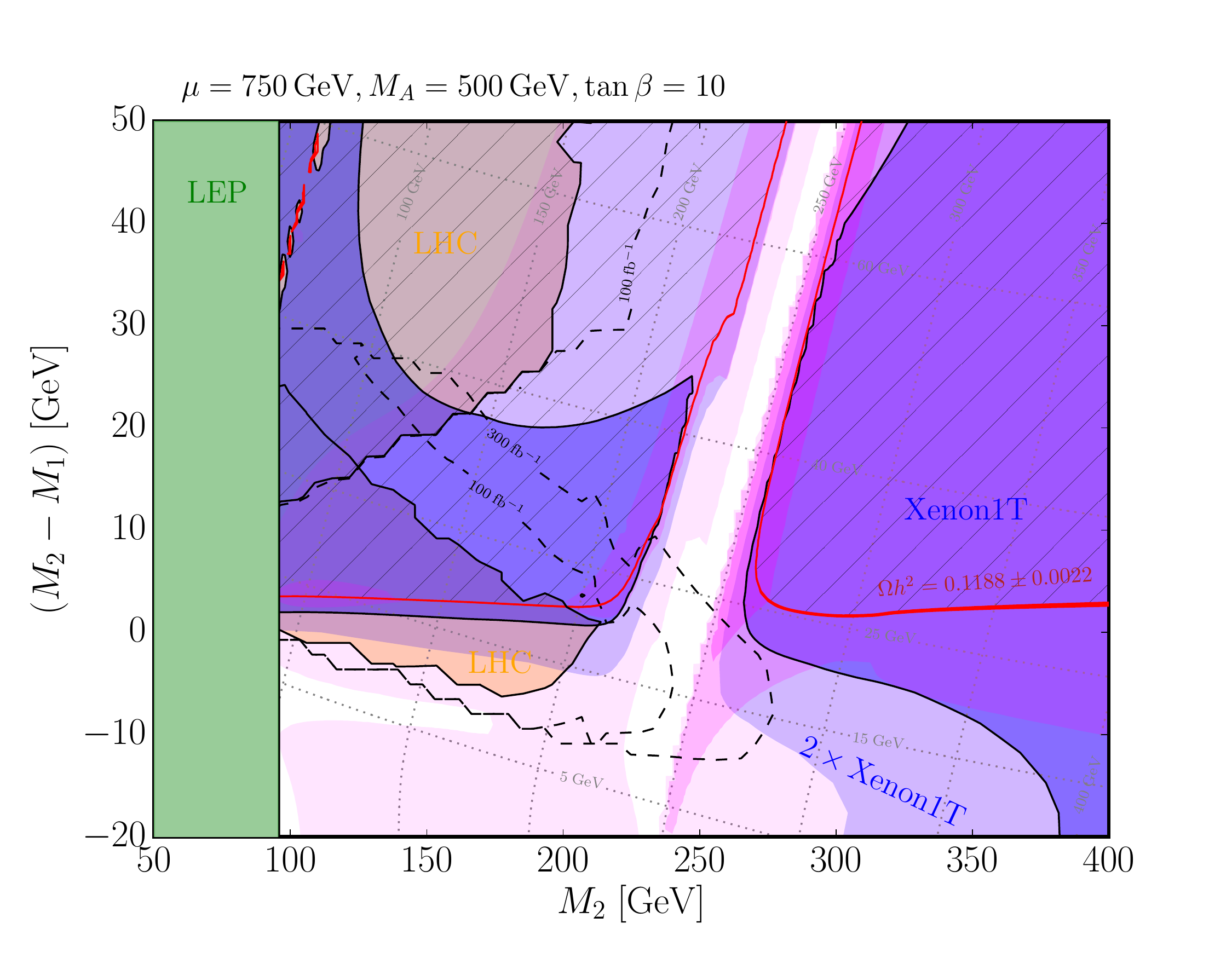}\hfill
\includegraphics[width=0.5\textwidth]{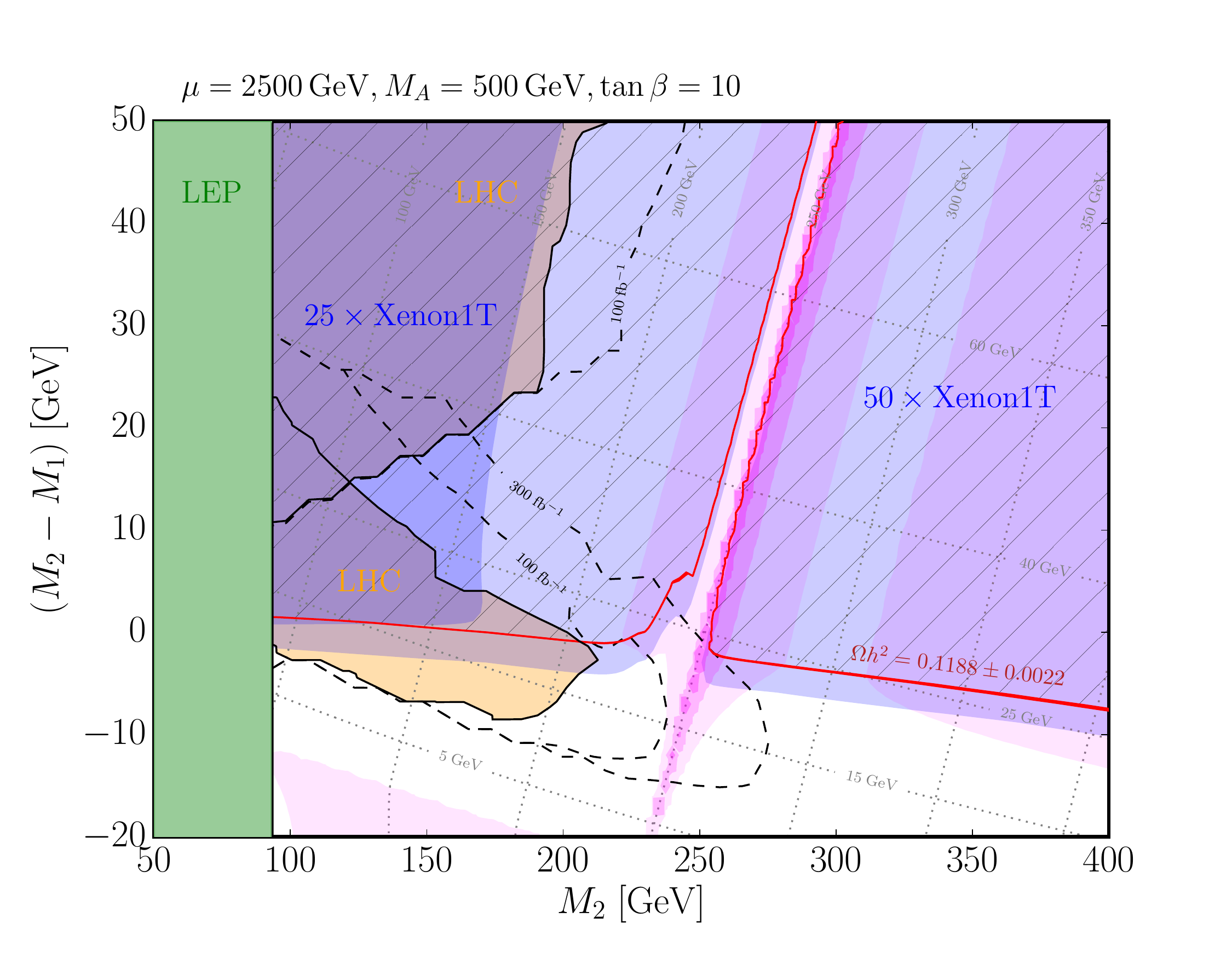}}
\caption{Comparison of current and projected constraints in the well-tempered bino-wino neutralino DM scenario in the ($M_2$, $M_2-M_1$) parameter plane for $\mu = 750\gev$ (\emph{left}) and $2.5\tev$ (\emph{right}), with $\tan\beta = 10$ and a pseudoscalar Higgs mass of $M_A = 500\gev$. The orange regions are excluded by two different LHC searches for $pp\to \chargpm{1}\neut{2} \to W^\pm Z\neut{1}\neut{1}$ by the CMS experiment~\cite{CMS:2017fij,CMS:2017fdz} (see text). Projected limits for $100\ifb$ and $300\ifb$ of the CMS searches are shown by dashed lines.
The exclusion obtained from the \LEP\ chargino mass limit is shown in green. Current and projected exclusions by the XENON1T DM direct detection experiment are shown as blue shaded areas with varying opacity (see blue text labels). Projected limits from DM indirect detection experiments are indicated by the magenta regions, where the current limit is scaled by factors of 10, 100 and 1000, with high to low opacity. The narrow region that predicts the observed DM relic density (including a $3\sigma$ error margin) is shown in red.
The roughly horizontal and vertical gray dotted contours give values for the mass splitting, $\Delta m = \mcharg{1} - \mneut{1}$, and the DM mass, $\mneut{1}$, respectively.
}
\label{fig:binowino}
\end{figure*}

We now move on to the bino-wino case, i.e. $M_1\simeq M_2\ll\mu$. Fig.~\ref{fig:binowino} shows current and future collider and dark matter constraints in the parameter plane defined by $M_2$ on the horizontal axis and the difference $(M_2-M_1)$ on the vertical axis. In the left and right panel we set $\mu = 750\gev$ and $2.5\tev$, respectively. We use the same values for $\tan\beta=10$ and $M_A=500$ GeV as in the bino-Higgsino case, and again choose the sfermion and gluino mass parameters to be $5\tev$. As before, the trilinear soft-breaking parameters are adjusted to yield a light Higgs boson mass of $M_h \simeq 125\gev$.

The hatched region features over-abundant relic dark matter, while the red stripes indicate the parameter region where the thermal relic density is within the range of the observed cosmological dark matter. Again, the orange regions are ruled out by current \LHC\ searches, and projections for $100\ifb$ and $300\ifb$ are given by the dashed lines. The exclusion from the \CMS\ search for compressed electroweakino mass spectra~\cite{CMS:2017fij} appears at small values of $(M_2-M_1) \sim -5$ to $20\gev$, while the other region at larger $(M_2-M_1)$ values (within the over-abundant relic DM region) is excluded by the multi-lepton plus $\etmiss$ search~\cite{CMS:2017fdz}. The green region is ruled out by \LEP\ chargino searches. In the left panel, the dark blue region is excluded by the current \XenonT\ direct detection results, while in the right panel, it shows the sensitivity if the current \XenonT\ limit is improved by a factor of 25. Furthermore, the light blue region shows future sensitivities for a rescaling of the \XenonT\ limit by a factor of 2 and 50 in the left and right panel, respectively. The magenta regions again indicate the sensitivity of indirect dark matter detection assuming an improvement by one, two and three orders of magnitude of the current \FermiLAT\ limit, going from high to low opacity.

The heavy Higgs funnel appears around the region where the lightest neutralino mass is close to $M_A/2=250\gev$, which naturally shows up as a diagonal strip on the [$M_2$, $(M_2-M_1)$] parameter plane. Apart from the funnel region, the difference $(M_2 - M_1)$ is required to be $\lesssim 5\gev$ and quite finely tuned in order to predict the observed DM relic abundance. Note also that for $M_2 < M_1$ the lightest neutralino is wino-like and vastly under-abundant for most of the mass values we focus on here.

The key observation from these results is that for sufficiently low values of $\mu$ direct detection experiments rule out large swaths, although not all, of the parameter space compatible with the correct thermal relic abundance. Again, the funnel region is problematic for direct detection searches due to the suppression of the detection rate with the neutralino thermal relic density. For larger $\mu$ values the situation is entirely different, and direct detection quickly becomes ineffective. As we noted earlier, the Higgsino-gaugino mixing drives the coupling of the neutralinos to the Higgs bosons. As large values of $|\mu| \gg M_1, M_2$ effectively decouple the gaugino-like neutralinos from the Higgs sector, the couplings between these neutralinos and the Higgs bosons become very small. Consequently, the rates for dark matter scattering off of ordinary matter are  highly suppressed at large $\mu$ values.

\begin{figure*}[t!]
\centering
\includegraphics[width=0.5\textwidth]{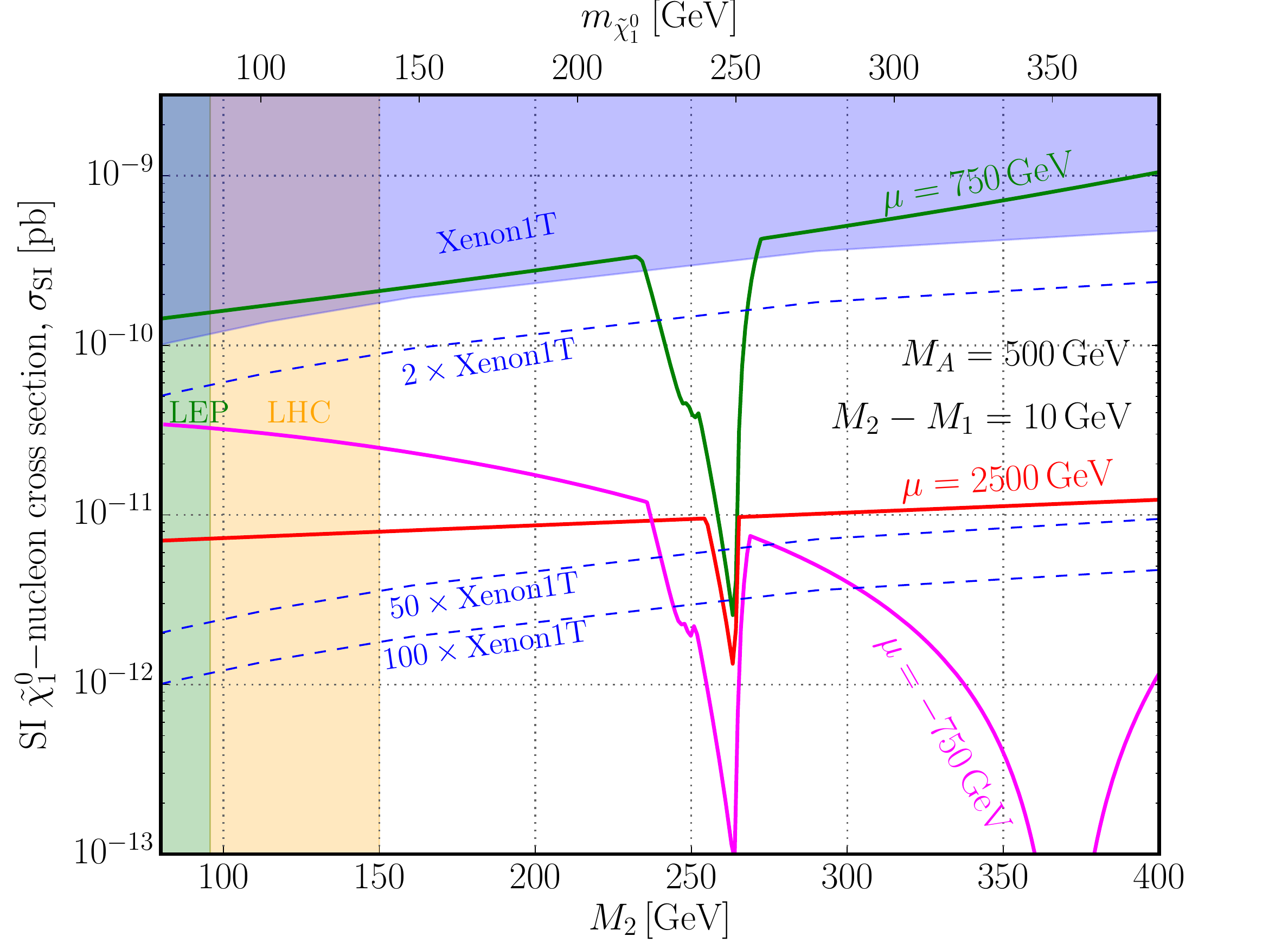}\hfill
\includegraphics[width=0.5\textwidth]{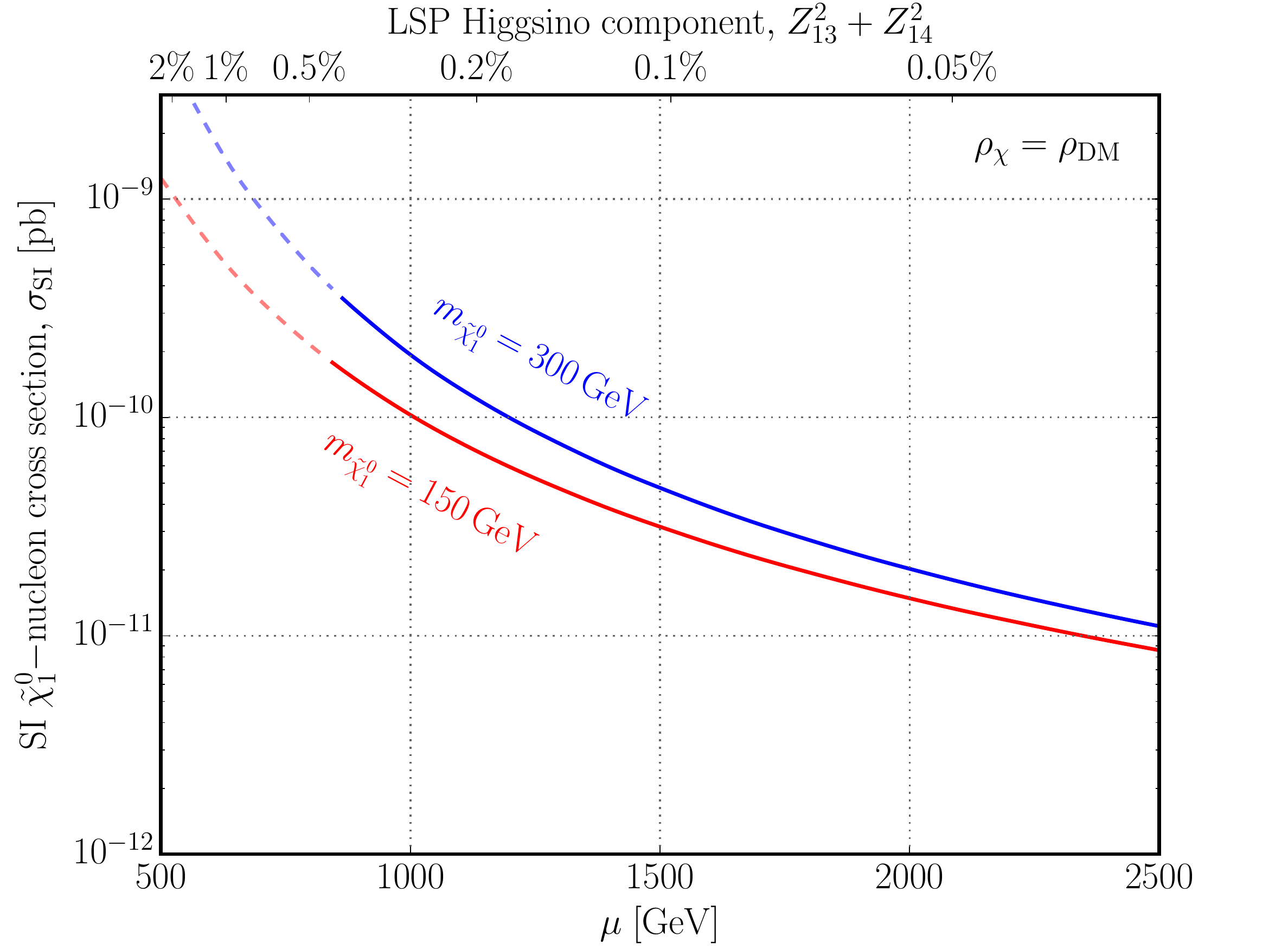}
\caption{Rescaled spin-independent (SI) neutralino-nucleon scattering cross section, $\sigma_\text{SI}$ (in $\mathrm{pb}$), in the bino-wino neutralino DM scenario. \emph{Left:} dependence on $M_2$ for fixed $M_1 = M_2 - 10\gev$, pseudoscalar Higgs mass $M_A =500\gev$, $\tan\beta = 10$ and values of $\mu=750\gev$ (\emph{green}), $ 2.5\tev$ (\emph{red}) and $-750\gev$ (\emph{magenta}). The second $x$-axis indicates the lightest neutralino mass, $\mneut{1}$ (assuming $\mu=750\gev$). The blue, green and orange filled regions are excluded by \LUX, \LEP\ and \LHC, respectively, whereas the dashed blue lines show different  rescalings of the current limit (\emph{see text labels}); \emph{Right}: dependence on $\mu$ for lightest neutralino masses of $\mneut{1} = 150\gev$ (\emph{red}) and $300\gev$ (\emph{blue}). The parameters are tuned such that the predicted DM relic density, $\rho_\chi$, matches its observed value, $\rho_\mathrm{DM}$. The second $x$-axis indicates the Higgsino component of the lightest neutralino, $Z_{13}^2 + Z_{14}^2$. In the dashed part of the lines $\sigma_\text{SI}$ is in conflict with the \LUX observations and thus excluded by \LUX.}
\label{fig:binowino_sigmaDD}
\end{figure*}

We further illustrate this feature in Fig.~\ref{fig:binowino_sigmaDD}. In the left panel we show the rescaled spin-independent neutralino DM-nucleon scattering cross section, $\sigma_{SI}$, for a slope along $M_2$ through our bino-wino benchmark scenario (Fig.~\ref{fig:binowino}) defined by $(M_2 - M_1) = 10\gev$, for the choices $\mu = 750\gev$ (\emph{green line}) and $2.5\tev$ (\emph{red line}), as well as for $\mu=-750\gev$ (\emph{magenta line}). The second $x$-axis gives the lightest neutralino mass corresponding to the $M_2$ values on the primary $x$-axis.
We also include the current \XenonT\ limit (\emph{blue region}) as well sensitivity curves for improvement factors of $2$, $50$ and $100$ of the current limit (\emph{blue dashed lines}). We can easily identify the funnel region, where the predicted neutralino relic density is lower than the observed dark matter relic abundance, $\xi < 1$, and $\sigma_{SI}$ therefore drops due to the rescaling with $\xi$. Moreover, we see that $\sigma_{SI}$ is more than one order of magnitude lower for $\mu=2.5\tev$ than for $\mu=750\gev$. In the case $\mu = -750\gev$, the spin-independent DM direct detection rates are again suppressed by destructive interference between the light and heavy CP-even Higgs exchange. This direct detection blind-spot appears around $\mneut{1}\sim 350\gev$, and shows an impact over a broad neutralino mass region. Hence, no exclusions arise from the current \XenonT\ limit for this scenario, and future results may or may not be sensitive (depending on how well the blind-spot is realized).

The right panel of Fig.~\ref{fig:binowino_sigmaDD} provides a different perspective onto this phenomenon. Here, we choose two values for the lightest neutralino mass outside the Higgs funnel region, $\mneut{1} = 150\gev$ (\emph{red line}) and $300\gev$ (\emph{blue line}), and show $\sigma_{SI}$ as a function of the Higgsino mass parameter $\mu$. Along these slopes, the parameters $M_1$ and $M_2$ are tuned in order to match both the chosen neutralino mass \emph{and} the observed dark matter relic density, $\xi = 1$. On the second $x$-axis we indicate the Higgsino component of the lightest neutralino, $Z_{13}^2 + Z_{14}^2$, corresponding to the $\mu$ values on the primary $x$-axis. The dashed parts of the lines at $\mu \lesssim 800\gev$ are excluded by the current \XenonT\ limit.
The plot illustrates clearly how $\sigma_{SI}$ decreases with the decreasing Higgsino component of the lightest neutralino, which in turn is realized by an increasing Higgsino mass parameter $|\mu| \gg M_1, M_2$.

\begin{figure*}[t]
\centering
\includegraphics[width=0.5\textwidth]{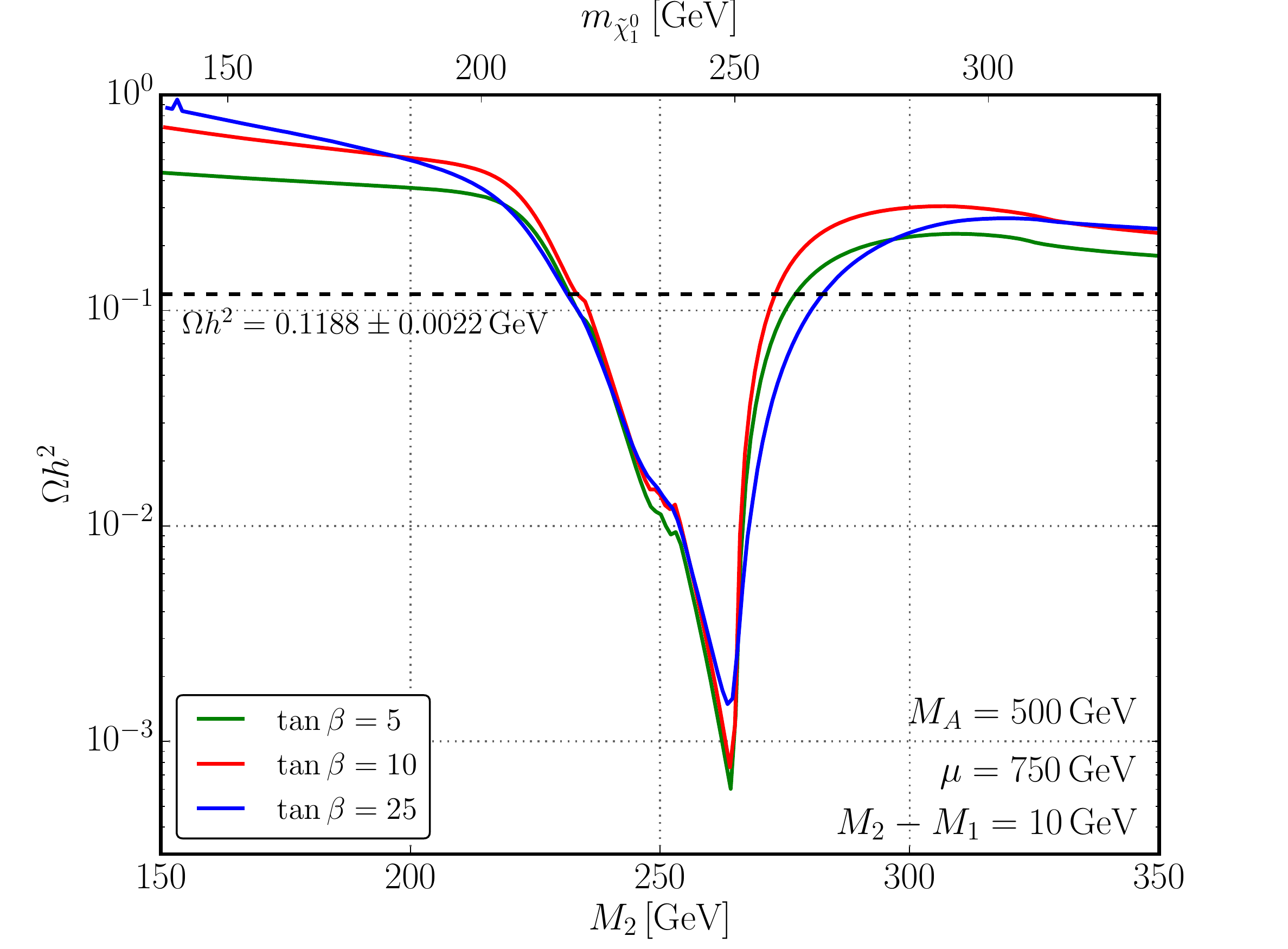}\hfill
\includegraphics[width=0.5\textwidth]{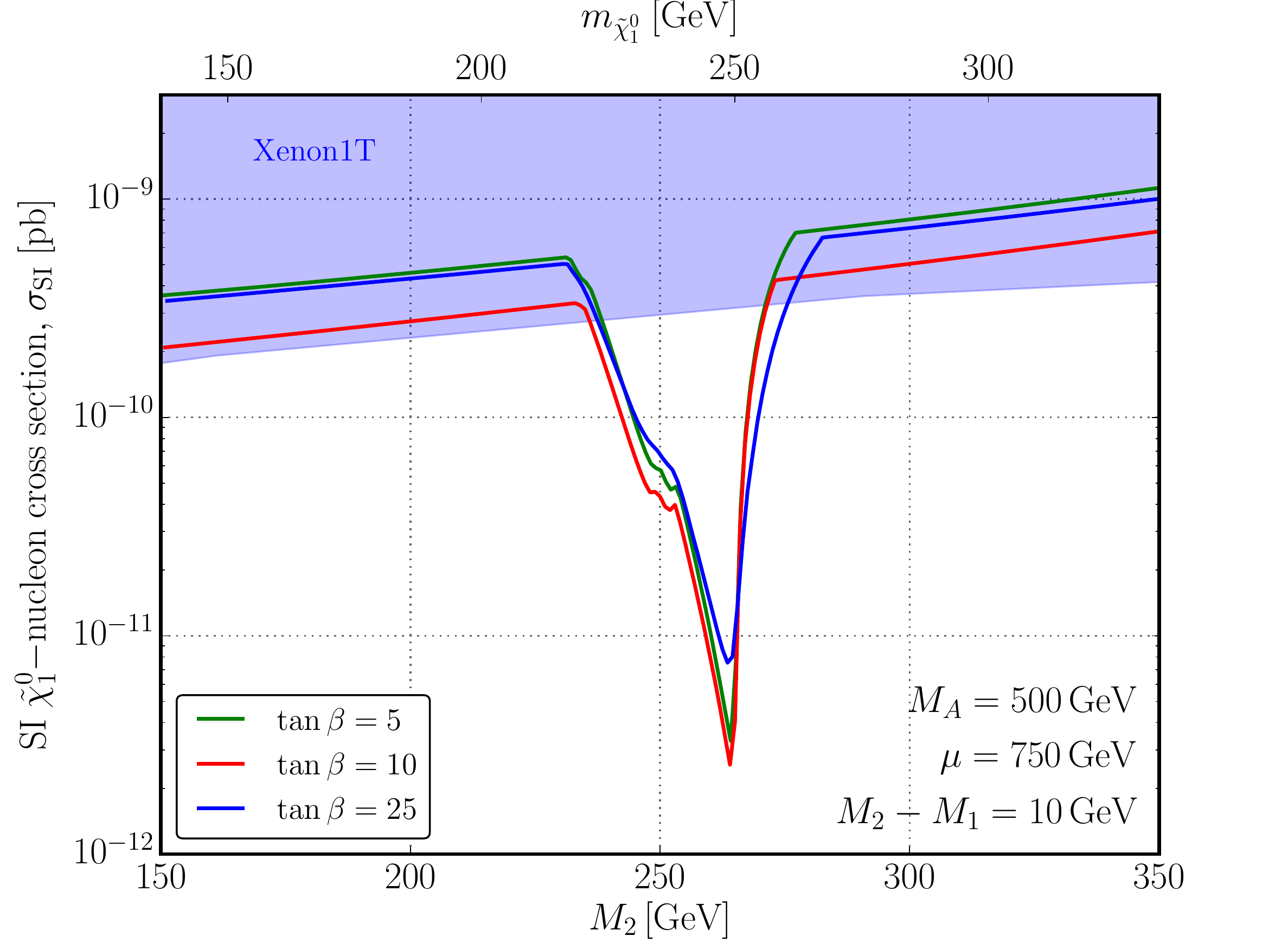}
\caption{Illustration of the Higgs funnel mechanism: neutralino DM relic abundance, $\Omega h^2$, (\emph{left}) and rescaled spin-independent neutralino DM-nucleon scattering cross section (\emph{right}) as a function of $M_2$, for fixed values of $\mu=750\gev$, $M_1 = M_2 - 10\gev$, a pseudoscalar Higgs mass of $M_A=500\gev$, and three values of $\tan\beta = 5,10$ and $25$. The upper horizontal axis gives the lightest neutralino mass, $\mneut{1}$, corresponding to the $M_2$ values on the lower horizontal axis (assuming $\tan\beta=10$).}
\label{fig:omega_funnel}
\end{figure*}

Let us briefly return to the discussion of Fig.~\ref{fig:binowino} and comment on the collider and indirect detection constraints.
LHC searches for direct electroweakino production are sensitive to significant portions of the bino-wino parameter space, including regions of correct thermal relic abundance. The current exclusion from the CMS search for compressed electroweakino mass spectra extends up to $\mneut{1}\sim 200\gev$. These constraints are largely independent of $\mu$ and form therefore important complementary probes of bino-wino dark matter besides direct detection experiments. In fact, for very large $\mu$ values, e.g.~for $\mu = 2.5\tev$ [Fig.~\ref{fig:binowino}(right)], these collider searches provide the \emph{only} relevant experimental constraint at the moment (besides the constraint from the observed DM relic abundance).\footnote{Note that for even larger $\mu$ values in the multi-TeV range the second neutralino can acquire a macroscopic decay length, which warrants dedicated displaced vertex searches at the LHC~\cite{Nagata:2015pra}.} Increased integrated luminosities of $100\ifb$ and $300\ifb$ will entail probing masses up to roughly $240\gev$ and $280\gev$, respectively. Furthermore, the best prospects for light, wino-like not-so-well tempered neutralinos (with $\xi <1$) lie in these collider searches.

As in the bino-Higgsino case, indirect detection is still ineffective at probing any of the shown parameter space. However, already with a factor of $10$ improvement of the current \FermiLAT\ limit significant portions of the Higgs funnel region with $\xi \lesssim 1$ become accessible, which are inaccessible to collider searches and, if $\mu$ is large, direct detection experiments in the foreseeable future.

Since in our analysis we focus for simplicity on one representative value of $\tan\beta$, we close this section with a comment on how the resonant pair-annihilation mechanism via the heavy Higgs bosons $H$ and $A$ depends on the choice of $\tan\beta$. In the left and right panels of Fig.~\ref{fig:omega_funnel} we show the thermal relic abundance, $\Omega h^2$, and the re-scaled spin-independent neutralino DM-nucleon scattering cross section, $\sigma_{SI}$, respectively, as a function of  $M_2$, for a fixed value of $\mu=750\ \gev$ and for a constant bino-wino mass difference, $(M_2 - M_1) = 10\gev$. 
As in Fig.~\ref{fig:binowino_sigmaDD}(left), we indicate on the upper horizontal axis the lightest neutralino mass, $\mneut{1}$, corresponding to the $M_2$ values on the lower horizontal axis (assuming $\tan\beta=10$). The red line corresponds to $\tan\beta=10$, our previous benchmark value, while the green and blue lines indicate, respectively, $\tan\beta=5$ and $25$. Both plots illustrate that while some variation with $\tan\beta$ exists, the qualitative picture is unchanged and our results obtained for $\tan\beta =10$ are thus broadly applicable to larger or smaller values of $\tan\beta$.

\section{Discussion and Conclusions}\label{sec:conclusions}

The standard lore is that the well-tempered neutralino is under siege, especially from recent results of direct dark matter detection experiments, and that collider searches for well-tempered neutralinos are generally outdone by direct searches for dark matter. Here, we considered {\em not-so-well tempered} neutralinos, defined as the ensemble of models both with the ``correct'' and with under-abundant thermal relic densities, and where we assume exclusively thermal production for under-abundant models  (thus implicitly invoking an additional dark matter species to explain the rest of the universal non-baryonic dark matter).

We focused on neutralino and chargino masses in a range potentially accessible to collider searches with the LHC, and systematically compared collider searches with direct and indirect searches for neutralino dark matter. In particular, we considered models with the lightest neutralino being a bino-Higgsino or bino-wino admixture, since both the limiting cases of pure bino, wino and Higgsino, and the wino-Higgsino case do not offer any interesting interplay between collider and dark matter searches.

We find that, as in the standard lore, bino-Higgsino well-tempered neutralinos are indeed ruled out by direct searches for sub-TeV neutralino masses, assuming effects from the MSSM Higgs sector can be neglected. In contrast, if the latter assumption does not hold, there can be two important effects, namely: (i) resonant annihilation through the neutral heavy Higgs bosons (``\emph{Higgs funnels}''), and (ii) suppression of the spin-independent DM-nucleon scattering cross section through destructive interference between processes mediated by the light and heavy CP-even Higgs boson (``\emph{blind spots}''). These blind spots can lead to a strong mitigation of the direct detection constraints in relevant regions of the parameter space where $M_1$ and $\mu$ have opposite sign. Such regions can then be probed by collider searches for electroweakinos (if they are light enough), which thus provide important complementary constraints. Here, dedicated LHC searches for compressed electroweakino mass spectra are particularly important, as they are sensitive to parameter regions that do not feature an over-abundant relic neutralino. In fact, in the bino-Higgsino case, these searches mostly probe the \emph{not-so-well tempered} neutralinos for which direct detection is flux-suppressed. This illustrates again the important complementarity of LHC searches and direct detection experiments.


In the bino-wino case, the relevance of direct detection constraints depends strongly on the value of $\mu$ with respect to $M_1$ and $M_2$. This is because the relevant neutralino coupling to a neutral Higgs boson depends on both the gaugino and Higgs\-ino fraction of the neutralino. Hence, in the limit where $|\mu|\gg M_1, M_2$, this coupling is suppressed. We find that, for $\mu \lesssim 800\gev$, the \XenonT\ limit is still able to exclude the well tempered bino-wino neutralino (with the correct thermal relic abundance). For larger $\mu$ values, however, current spin-independent direct detection limits do not yield any constraints on the parameter space. Note also, that in the case $\mu<0$, blind spots for dark matter direct detection searches can again appear due to the cancellation of the contributions from light and heavy CP-even Higgs exchange, and thus no robust statement about the \XenonT\ constraints on $\mu$ can be made in this case. 
In contrast, collider searches for electroweakinos are largely independent of $\mu$ (as long as $|\mu| > M_1, M_2$), and currently constitute the \emph{only} direct probe if $|\mu| > 800\gev$. Again, searches focusing on compressed electroweakino mass spectra are particularly important, as they are sensitive to both scenarios with well-tempered and not-so-well tempered neutralino dark matter. Current search results from the CMS collaboration exclude neutralino masses up to $\sim 200\gev$, with prospective sensitivity to mass values of $240\gev$ and $280\gev$  with integrated luminosities of $100\ifb$ and $300\ifb$, respectively.


We find that in all cases indirect searches via gamma-ray observations do not offer competitive constraints with respect to direct detection experiments and collider searches for the not-so-well tempered neutralino.

In conclusion, LHC searches for electroweakinos continue to be well-motivated in the context of supersymmetric models with heavy sfermions and gluinos. In particular, LHC searches are highly complementary to dark matter direct detection experiments in the case of ``not-so-well-tempered'' neutralino models, i.e.\ models that feature an under-abundant thermal relic density, as well as in models with a bino-wino mixed lightest neutralino. In fact, in the latter case, for large $\mu$ values, collider searches are currently the only means to directly probe these scenarios.

\section*{Acknowledgments}
We thank Marthijn Sunder for helpful assistance with \texttt{Resummino}. We are grateful to David Curtin who courteously provided a helpful script for digitizing color plots.
We furthermore thank Roberto Castello and Lesya Shchutska for helpful comments on Refs.~\cite{CMS:2016zvj,CMS:2016gvu}. This work is partly supported by the US Department of Energy, grant number DE-SC0010107. TS furthermore acknowledges support by the Alexander von Humboldt foundation through a Feodor-Lynen research fellowship.

\bibliographystyle{apsrev4-1}
\bibliography{main}
\end{document}